\documentclass[journal,twoside]{IEEEtran}

\usepackage{chemarrow}
\usepackage{CJK}
\usepackage{fancybox}
\usepackage{times}
\usepackage{amsmath}
\usepackage{amssymb}
\usepackage{graphicx}
\usepackage{subfigure}
\usepackage{bm}
\usepackage{cite}

\usepackage{array}
\usepackage[linesnumbered,ruled]{algorithm2e}

\usepackage{algorithmic}

\usepackage{amsfonts,amssymb,amsmath,bm}
\usepackage{amsthm}
\usepackage{multicol}
\usepackage{graphicx,subfigure}
\usepackage{cite,color}
\usepackage{booktabs}
\usepackage{threeparttable}
\usepackage{mathtools}
\usepackage{verbatim}
\usepackage{subeqnarray}
\usepackage{cases}

\begin{document}

\title{Cross-Layer Control for Worse Case Delay Guarantees in Heterogeneous Powered Wireless Sensor Network via Lyapunov Optimization}

\author{Weiqiang Xu,~\IEEEmembership{Senior Member,~IEEE},
Jianchen Tu,
Qingjiang Shi,~\IEEEmembership{Member,~IEEE},

\thanks{Weiqiang Xu, Jianchen Tu, and Qingjiang Shi are with School of Information Science \& Technology, Zhejiang Sci-Tech University, Hangzhou, 310018, P. R. China. Email: wqxu@zstu.edu.cn, jianchentu@zstu.edu.cn, shiqj@zstu.edu.cn}
}

\maketitle

\begin{abstract}
The delay guarantee is a challenge in wireless sensor networks (WSNs), where energy constraints must be considered.
The coexistence of renewable
energy and electricity grid
is expected as a promising energy supply manner for WSNs to
remain function for a potentially infinite lifetime.
In this paper, we address cross-layer control to guarantee worse case delay for Heterogeneous Powered (HP) WSNs.
We design a novel
virtual delay queue structure, and apply the Lyapunov optimization
technique  to develop cross-layer control algorithm only requiring knowledge of the instantaneous system state, which provides efficient throughput-utility,
and guarantees bounded worst-case delay.
We analyze the performance of the proposed algorithm and verify the theoretic claims
through the simulation results.
Compared to the existing work, the algorithm presented in this
paper achieves much higher optimal objective value with ultra-low data drop due to the proposed novel virtual queue structure.
\end{abstract}


\section{Introduction}
Wireless sensor networks (WSNs) have been an active research area during the last two decades.
By embedding low-cost, low-power, small-size, and multifunctional sensor nodes into the environment,
a variety of parameters such as pressure, humidity, temperature, and vibration intensity are measured and wirelessly transmitted to a processing center. Based on these collected parameters, the center then analyzes any potential problems, and even rapidly responds to real-time events with appropriate actions.
Due to self-organization, rapid deployment, easy maintenance, and reduced cost, WSNs provides several potential advantages over traditional wired system. The existing and potential applications of WSNs span a very wide range, including industrial monitoring and control\cite{Gungor_Hancke2009}, building automation\cite{Cao_Chen2010},  video surveillance\cite{Wu2011}, and so on.

Traditionally,
sensor nodes are powered by a non-rechargeable battery with limited energy storage capacities.
Thus, the main research efforts in developing WSNs have focused on how to
improve the energy efficiency with respect to limited battery energy\cite{Mo2011,Wang2013}.
 Recently, energy harvesting (EH) technique utilized in wireless system has attracted attention\cite{Sudevalayam_Survey}.
However, due to the low recharging rate and the time-varying profile
of the energy replenishment process, renewable
energy cannot guarantee to provide the perpetual operation for WSNs in most application scenarios.
The coexistence of renewable
energy and electricity grid, called as Heterogeneous Power (HP) in this paper,
is expected as a promising energy supply manner to
remain function for a potentially infinite lifetime in wireless system\cite{Gong_Niu2013}.

There are a number of different real-time requirements in WSNs applications,
for instance, manufacturing and process automation and motion control.
Thus,
delay constrained data collection in WSNs has
been studied to some extent\cite{Hariharan2013,Zheng2014}.
However, there are only few works for diminishing delay in
EH-WSNs, which are mainly based on duty cycle adjustment\cite{Yoo_Shim2012}.
However, these works do not investigate to reduce queueing delay with deterministic or probabilistic guarantee.
In a multihop network, the network delay performance depends heavily on the queue length at every node along the multihop route.
When the data arrive at a node, they have to be processed and forwarded.
If the data arrive faster than the node can process them, the node puts them into the queue until it can get around to transmit them.
As a queue begins to fill up due to the traffic arriving faster than it can be processed, the amount of delay a packet experiences going through the queue increases. The longer is the line of data waiting to be transmitted, the longer is the average queueing delay. In practice, each node only has a finite buffer to hold the data. Thus, a node may experience a full queue that may potentially cause the loss of data traffic, leading to QoS degradation.
Although the queueing delay have been extensively addressed in multi-hop wireless networks (See the related works in Section II).
However,
these works are not readily extendable to WSNs, where energy constraints must be considered.
Furthermore, most of existing
works provide only
an average delay bound, can not give bounds on the delays
of individual sessions, and even yield unbounded worst-case delays.
It is vital to propose a scheme to guarantee worst-case delay for WSNs in a variety of  applications.

In this paper, we will address cross-layer control to guarantee worse case delay for HP-WSNs.
through designing a novel
virtual delay queue structure, and applying the Lyapunov optimization technique.
The key contributions of this paper are as follows.

\begin{itemize}
  \item We consider heterogeneous energy supplies from renewable
energy, electricity grid and mixed energy, multiple energy consumptions due to sensing, transmission and reception,
and multi-dimension stochastic natures due to EH profile, electricity price and channel condition.
We develop a novel virtual delay queue scheme to share the burden of actual
packet queue backlogs to guarantee specific delay performances and finite data buffer sizes.
Finally, we formulate a discrete-time stochastic delay-aware cross-layer control problem for achieving the optimal trade-off between the time-average throughput utility and electricity cost subject to the data and energy queuing stability constraints, and to guarantee worse case delay for HP-WSNs.

  \item To obtain a distributed and
low-complexity solution,
we apply the Lyapunov drift-plus-penalty
technique \cite{Neely2006Book} to transform the stochastic control problem into a deterministic optimization problem, which can be solved by a greedy algorithm at every time-slot only requiring knowledge of the instantaneous system state.
Furthermore,
by exploiting the special structure of the deterministic optimization problem, we design a distributed algorithm which decomposes the overall problem into the energy management subproblem, the physical layer subproblem, the network layer subproblem, and the transport layer subproblem.

  \item We analyze the performance of the proposed distributed algorithm.  We show that a control parameter $V$  enables an explicit trade-off between the average utility and queue backlog. Specifically, the proposed distributed algorithm is shown only achieve a time average utility that is within ${\cal {O}}(1/V)$ of the optimal network utility for any $V \ge 0$, while ensuring that the average network backlog is ${\cal  {O}}(V)$,  when the system state is independent and identically distributed (i.i.d.).
Finally, extensive simulations verify the theoretic claims, and demonstrate that the proposed distributed algorithm achieves much higher optimal objective value with ultra-low data drop, compared to the existing work.
\end{itemize}


Throughout this paper, we use the following notations.
The probability of
an event $A$ is denoted by Pr$(A)$. For a random variable
$X$, its expected value is denoted by $\mathbb{E}[X]$ and its expected
value conditioned on event $A$ is denoted by $\mathbb{E}[X|{A}]$.
The indicator function for an event $A$ is denoted by $\textbf{1}_{A}$; it equals
1 if $A$ occurs and is 0 otherwise. ${{{[x]}^ + } = \max (x,0)}$.

The remainder of the paper is organized as follows. In Section II, we introduce the related works. In Section III, we give the system model and problem formulation.
In Section IV, we present
the distributed cross-layer optimization algorithm using Lyapunov optimization.
In Section V, we present the performance analysis of
our proposed algorithm.
Simulation results are given in Section VI. Concluding remarks are provided in Section VII.

\section{Related Work}

\subsection{Optimization for EH node and EH-WSNs}

Recently, a great deal of research efforts have been devoted to investigate the
energy management and data transmission schemes in EH node.
Sharma et al. in \cite{Sharma2010} obtain two energy management policies to achieve the optimal throughput and the minimal mean delay.
Srivastava et al. in \cite{Srivastava2013} give an energy management scheme to achieve the optimal utility asymptotically while keeping both the battery discharge and data loss probabilities low.
Zhang et al. in \cite{ZhangTWC2013} address the adaptive decision of the sampling rate for EH sensor node with a limited battery capacity to maximize the overall network performance.
Mao et al. in \cite{MaoTVT2014} study the energy allocation for sensing and transmission in an EH sensor node with a rechargeable battery and a finite data buffer.
However, different nodes in networks may have quite different workload requirements and available energy sources.
Some works address the optimal design for EH-WSNs.
Mao et al. in \cite{Mao2012TAC} address the joint control of the data queue and battery buffer to maximize the long-term average sensing rate of  EH-WSNs under certain QoS constraints for the data and battery queues.
Chen et al. in \cite{Chen2012INFOCOM} address the joint problem of energy allocation and routing to maximize the total system utility, without prior knowledge of the replenishment profile.
Sarkar et al. in \cite{Sarkar2013} design routing and scheduling policies that optimize network throughput in EH-WSNs.
Gatzianas et al. in \cite{Gatzianas2010} design
an online adaptive transmission scheme to achieve close-to-optimal utility performance and to ensure the data queue stability for wireless networks with rechargeable battery.
Huang et al. in \cite{Huang_Neely2013} develop the Energy-limited Scheduling Algorithm (ESA) and Modified-ESA (MESA) algorithm to achieve
an explicit and controllable tradeoff between optimality gap and queue sizes for EH-WSNs.
Tapparello et al. in \cite{Tapparello2014}
proposed the joint optimization scheme of source coding and transmission to minimize the reconstruction distortion cost for EH-WSNs with the correlated sources measurement.
However, delay metrics is not considered in the works mentioned above.
Furthermore, almost no works, except\cite{MaoTVT2014}\cite{Tapparello2014}, study the joint energy allocation for communication module and sensing module together.
In addition,
due to the low recharging rate and the time-varying profile
of the energy replenishment process, sensor nodes solely powered by harvested energy can not guarantee to provide reliable services for the perpetual operation.

\subsection{Delay-aware optimization for multi-hop wireless networks}

The optimization-based design methodology has been extensively developed to handle the queueing delay in multi-hop wireless networks\cite{survey_yingcui}.
In \cite{Gupta_Shroff_infocom2009}, Gupta et al. analyze the delay performance of a multihop wireless network with a fixed route and arbitrary interference constraints.
In \cite{Venkataramanan_Lin2010}, Venkataramanan et al. derive bounds on the best performance of end-to-end buffer usage over a network with general topology and with fixed, loop-free routes.
In \cite{Bui_Srikant2009}, Bui et al. propose a novel architecture and algorithm to improve the delay performance of the back-pressure algorithm.
In \cite{Ying_Shakkottai2009}, Ying et al. propose a hop-count based queueing structure to adaptively select a set of optimal
routes based on shortest-path information. resulting in much smaller end-to-end packet delays as compared to the traditional back-pressure algorithm.
In \cite{Xiong_Li2011}, Xiong et al. propose a novel link rate allocation strategy and a regulated scheduling strategy to develop delay-aware joint flow control, routing, and scheduling algorithm to achieve loop-free route and optimal network utilization for general multi-hop networks.
However, none of the above-mentioned works provides explicit end-to-end delay guarantees. There are several works aiming to address end-to-end delay or buffer occupancy guarantees in multihop wireless networks.
In \cite{Huang_Lin2013}, Huang et al. proposes a fully-distributed joint congestion control and scheduling algorithm that can guarantee order optimal per-flow end-to-end delay and utilize close-to-half of the system capacity for multihop wireless networks with fixed-routing under the one-hop interference constraint.
In \cite{Le_Modiano2012}, Le et. al. investigate the performance of joint flow control, routing, and scheduling algorithms that achieve high
network utility and deterministically bounded backlogs in wireless networks with finite buffers.
In \cite{Xue_Ekici2013}, Xue et al. propose a joint congestion control,
routing, and scheduling problem in a multihop wireless network
to satisfying per-flow average end-to-end delay constraints and
minimum data rate requirements.
Note that the delay threshold is a time-averaged
upper bound, not a deterministic one.
These prior works may yield unbounded worst-case delays.
In \cite{Neely_opportunistic2011}, Neely et al. design an opportunistic scheduling algorithm that guarantees all sessions have a bounded worst case delay.
However,
these works are not readily extendable to multihop WSNs, where energy constraints must be considered. It is more challenge to deal with queue delay in EH-WSNs.

\section{System Model and Problem Formulation}
\label{sec:model}

Consider a multi-hop WSNs with $N$ nodes that operates in discrete time with normalized time slots
$t \in {\cal T}=\left\{ {0,1,...,T} \right\}$. Let
${\cal N} = {{\cal N}_H} \cup {{\cal N}_G} \cup {{\cal N}_M} = \left\{ {1,...,N} \right\}$
denote the set of sensor nodes in the network. ${\mathcal{N}_H}$ is the set of nodes powered by EH, called EH nodes,
${\mathcal{N}_G}$ is the set of nodes powered by electricity grid (EG), called EG nodes,
${\mathcal{N}_M}$ is the set of Mixed energy (ME) nodes powered by both EH and EG.
$\mathcal{L} = \left\{ {\left( {m,n} \right),m,n \in \mathcal{N}} \right\}$
represents the set of communication links.
Each node has multiple sensor interfaces and can measure multiple information.
We assume that there are $F$ traffic sessions, which can be measured by source nodes
$n \in {\mathcal{N}_s}$, ${\mathcal{N}_s} \subset \mathcal{N}$. Let
$\mathcal{F} = \left\{ {1,...,F} \right\}$
denote the set of traffic sessions in the network.

\subsection{Data Queue Dynamic}

The data backlog for each session $f \in \mathcal F$ and each node $n \in \mathcal N$ in slot $t$ is denoted by
$Q_{_n}^f\left( t \right) $.
The queue dynamics is given by:
\begin{eqnarray} \label{data_q_def}
  Q_n^f\left( {t + 1} \right) \leq \left[ {Q_n^f\left( t \right)} \right. - \sum\limits_{b \in \mathcal{O}\left( n \right)} {\mu _{nb}^f\left( t \right)}  - {\left. {D_n^f\left( t \right)} \right]^ + } \hfill \nonumber \\
  \quad \quad \quad \quad  + \sum\limits_{a \in \mathcal{I}\left( n \right)} {\mu _{an}^f}  + {\mathbf{1}}_n^fr_n^f\left( t \right) \hfill
\end{eqnarray}
Where ${\left[ a \right]^ + }$ denotes $\max \left\{ {a,0} \right\}$, $\mathcal{O}\left( n \right)$ denotes the set of nodes $b$ with
$\left( {n,b} \right) \in \mathcal{L}$. $\mathcal{I}\left( n \right)$ denotes the set of nodes $m$ with $\left( {m,n} \right) \in \mathcal{L}$. The service decision variable $\mu _{nb}^f\left( t \right)$ represents the amount of packets of session $f$ successfully served from node $n$ to node $b$ on slot $t$. The drop decision variable $D_n^f\left( t \right)$ represents the number of packets of session $f$ that dropped by node $n$ on slot $t$. The admission decision variable $r_n^f\left( t \right)$ represents the amount of packets of session $f$ that sensed by node $n \in \mathcal{N}_s$ on slot $t$.
We assume that the drop decision $D_n^f\left( t \right)$ and service decision $\mu _{nb}^f\left( t \right)$
are made at the beginning of each slot, and the admission decision $r_n^f\left( t \right)$ will be made at the end of each slot.
There exist a maximum transmission rate $\mu _{\max }$ over any link and a maximum amount $D_{\max }$ of data we are allowed to drop on slot $t$ for any session, which are both finite constants.
Then the drop decisions and service decisions will be subjected to the constraint $D_n^f\left( t \right) \leqslant {D_{\max }}$
and $\sum\limits_{f \in \mathcal{F}} {\mu _{n{\text{b}}}^f\left( t \right)}  \leqslant {\mu _{\max }}$,
respectively.
The admission decisions $r_n^f\left( t \right)$ is also subjected to the constraint
${r_n^f}\left( t \right) \leqslant {R_{\max }}$, where $R_{\max }$ is a finite constant.

\subsection{A Novel Virtual Queue Structure}

Now we consider the delay of data in the data queue. \cite{NeelyBook2010} developed a virtual queue called $\epsilon -persistent \; service \; queue$ for each node. Here we propose a novel virtual queue structure, called as the delay queue,  to guarantee the worst case delay of all sessions.

For each node $n \in \cal N$ and for each session  $f \in \mathcal{F}$, we define a virtual queue $\tilde Q_n^f\left( t \right)$ with the queue dynamics as follows:
\begin{subnumcases}
{\tilde Q_n^f\left( {t + 1} \right) =  \\}
\tilde Q_n^f\left( t \right) - \sum\limits_{b \in {\mathcal O}\left( n \right)} {\mu _{n{\text{b}}}^f\left( t \right)}  - D_n^f\left( t \right) + \varepsilon _n^f, \nonumber\\
\quad \quad \quad \quad \quad\quad \quad \quad  Q_n^f\left( t \right) > \rho\tilde Q_n^f\left( t \right) \label{qbar_queue_def_1} \\
  \tilde Q_n^f\left( t \right) - \mu _{\max }^{out} - {D_{\max }} + \varepsilon _n^f, \nonumber\\
  \quad \quad \quad \quad \quad\quad \quad \quad  Q_n^f\left( t \right) \leqslant \rho\tilde Q_n^f\left( t \right) \label{qbar_queue_def_2}
\end{subnumcases}
Where $\tilde Q_n^f\left( 0 \right){\text{ = }}0$, and
$\varepsilon _n^f$ is a constant satisfied with $0 < \varepsilon _n^f \leqslant {D_{\max }}$.
Shown in Eq. \eqref{qbar_queue_def_1}\eqref{qbar_queue_def_2},
there always exists a persistent arrival with the size $\varepsilon _n^f$ to the virtual queue at each slot.

If there exists a scheduling algorithm that maintains bounded $Q_n^f\left( t \right)$ and $\tilde Q_n^f\left( t \right)$, the worst case delay will also be bounded,  which will be proved in Theorem 1.
\newtheorem{0}{Theorem}
\begin{0}\label{lemma_1}
For all slots $t \in \cal T$ and traffic sessions $f \in \mathcal{F}$, suppose a scheduling algorithm is used to ensure that the queue $Q_n^f\left( t \right)$ and $\tilde Q_n^f\left( t \right)$ have the finite upper bound
${Q_{\max }}$ and ${\tilde Q_{\max }}$, respectively. Assuming First Input First Output (FIFO) service, then the worst case delay of all non-dropped data in node $n$ can be defined as $W_{n,\max }^f$, which is given by:
\begin{eqnarray}
W_{n,\max }^f = \max \left\{ \left[\left( 1+ \rho\right)Q_{\max }+ \rho\tilde Q_{\max}\right]/\left(\rho\varepsilon _n^f\right), \right.\nonumber\\
\left.2{\tilde Q_{\max }}/\left( {\mu _{\max }^{out} + D_{\max } - \varepsilon _n^f} \right) \right\} \label{worst_delay_bound}
\end{eqnarray}
\end{0}
\textbf{Proof}: Please see Appendix A.

To achieve the minimal worst case delay $W_{_{\max }}^f$ in node $n$, according to \eqref{worst_delay_bound}, we can see the optimal value of $\rho^*$ should be set
\begin{equation}\label{Optimal_rho}
{\rho ^*}{\rm{ = }}\frac{{{Q_{\max }}\left( {{\mu _{\max }^{out}} + {D_{\max }} - \varepsilon _n^f} \right)}}{{{2{\tilde Q}_{\max }}\varepsilon _n^f{\rm{ - }}\left({Q_{\max }+{\tilde Q}_{\max }}\right)\left( {{\mu _{\max }} + {D_{\max }} - \varepsilon _n^f} \right)}}
\end{equation}


\subsection{Data sensing/processing}

At time slot $t$, node $n$ will measure information source ${F_n}$ independently, where ${F_n} \in {\mathcal{F}_n}$. ${\mathcal{F}_n}$ is the set of information source, ${\mathcal{F}_n} = \left\{ {1,2,3,...,{F_n}} \right\}$ and $\mathcal{F} = \bigcup\limits_{n \in {\mathcal{N}_s}} {{\mathcal{F}_n}} $. The measured samples of the session ${F_n}$ is compressed with rate
$r_n^f\left( t \right)$ . We define $p_f^S\left( {r_n^f\left( t \right)} \right)$ as the function of energy consumption for sensing/processing at a particular rate $r_n^f\left( t \right)$. The relationship between $p_f^S\left( {r_n^f\left( t \right)} \right)$ and $r_n^f\left( t \right)$ can be regarded as linearity, i.e. $p_f^S\left( {r_n^f\left( t \right)} \right) = \tilde p_f^Sr_n^f\left( t \right)$ \cite{Tapparello2014}. $\tilde p_f^S$ denotes the energy consumption for sensing/processing per unit data of the $f$-th session. Similarly, we use $\tilde p_f^R$ to denote the energy consumption for node $n$ to receive one data from the neighbor nodes in the network.

\subsection{Data transmission}

We define the transmission power allocation matrix for data transmission at slot $t$ as below:
\[\mathbf p^T \left( t \right) = \left( p_{mn}^T\left( t \right),\left( {m,n} \right) \in \mathcal{L} \right)\]
$p_{mn}^T\left( t \right)$ denotes the transmission power allocated to link $\left( {m,n} \right)$ at slot $t$. Then for each node $n$, the power consumption should follow below condition:
\begin{equation}\label{p_t_constraint}
\sum\limits_{b \in \mathcal{O}\left( n \right)} {p_{nm}^T\left( t \right)}  \leqslant P_n^{\max },n \in \mathcal{N} \end{equation}
$P_n^{\max }$ is the maximal transmission power limitation at node $n$, which is assumed to be a finite constant.

For WSNs, there always exist interference between different links while transmission. We denote the signal to interference plus noise ratio (SINR) of link $\left( {n,b} \right)$ as the function of transmission power $\mathbf p^T\left( t \right)$ and network channel state $\mathbf S^C\left( t \right)$:
\begin{eqnarray}\label{SINR_def}
  {\gamma _{nb}}\left( t \right) &\triangleq& {\gamma _{nb}}\left( {\mathbf p^T\left( t \right),\mathbf S^C\left( t \right)} \right) \nonumber \\
  \quad \quad \;\;&=& \frac{{S_{nb}^C\left( t \right)p_{nb}^T\left( t \right)}}{{N_0^b + \sum\nolimits_{a \in {\mathcal{J}_{n,b}}} {\sum\nolimits_{\left( {a,m} \right) \in \mathcal{L}} {S_{ab}^C\left( t \right)p_{am}^T\left( t \right)} } }} \hfill
\end{eqnarray}
Where $\mathbf S^C\left( t \right)$ present the network channel state matrix, $\mathbf S^C\left( t \right) = \left\{ {S_{nm}^C\left( t \right),\left( {n,m} \right) \in \mathcal{L}} \right\}$. $S_{nb}^C\left( t \right)$ denotes the link fading coefficient on link $\left( {n,b} \right)$ at slot $t$,  which is randomly varying over time slots in an i.i.d. fashion according
to a potentially unknown distribution and taking non-negative values from a
finite but arbitrarily large set ${\cal S}^C$.
$N_0^b$ presents the background noise power at node $b$, ${\mathcal{J}_{n,b}}$ is the set of nodes whose transmission may interfere to the link $\left( {n,b} \right)$, excluding node $n$.

Furthermore, we define ${C_{nb}}\left( t \right) = \log \left( {1 + {\gamma _{nb}}\left( t \right)} \right)$ as the link capacity. So we can get that the data transmission constraint condition:
\begin{equation}
\sum\limits_{f \in \mathcal{F}} {\mu _{nb}^f\left( t \right) \leqslant {C_{nb}}\left( t \right)} \quad \forall n \in \mathcal{N},\forall b \in \mathcal{O}\left( n \right)
\label{miu_constraint1}
\end{equation}

In the high-SINR case, $\log \left( {{\gamma _{nb}}\left( t \right)} \right)$ would have been a good approximation of
$\log \left( {1 + {\gamma _{nb}}\left( t \right)} \right)$. Thus, we will regard ${\tilde C_{nb}}\left( t \right) = \log \left( {{\gamma _{nb}}\left( t \right)} \right)$ as the link capacity in the following context. So the constraint \eqref{miu_constraint1} can be transformed into:
\begin{equation}\label{miu_constraint2}
\sum\limits_{f \in \mathcal{F}} {\mu _{nb}^f\left( t \right) \leqslant {{\tilde C}_{nb}}\left( t \right)} ,\forall n \in \mathcal{N},\forall b \in \mathcal{O}\left( n \right)
\end{equation}

\subsection{Energy Consumption Model}

According to the above description, we can get the total energy consumption of node $n$ at slot $t$ to accomplish the tasks, including data sensing/processing, data transmission and data reception:
\begin{eqnarray}\label{p_consumption_def}
p_n^{Total}\left( t \right) = \sum\limits_{f \in \mathcal{F}} {\tilde p_f^S{r_n^f}\left( t \right)}  + \sum\limits_{b \in \mathcal{O}\left( n \right)} {p_{nb}^T\left( t \right)} \nonumber \\
+ \tilde p_n^R\sum\limits_{a \in \mathcal{I}\left( n \right)} {\sum\limits_{f \in \mathcal{F}} {\mu _{an}^f\left( t \right)} }
\end{eqnarray}

\subsection{Energy Queue Dynamic}

We define ${E_n}\left( t \right)$ as the energy queue in node $n$ at slot $t$. The EH nodes can harvest energy ${e_n}\left( t \right)$ from the  environment (such as sunshine), then store the energy into the battery. The energy in EG nodes is generally acquired from the electricity grid. Similarly for the EH nodes, the energy in the EG nodes will also be stored into the battery. The energy supplied by the electricity grid is denoted as ${g_n}\left( t \right)$.
Different from the EH nodes and EG nodes, the ME nodes can both harvest energy from environment and acquire energy from the electricity grid.

Then we can give the energy queue dynamic for any node $n$ in the network as follows:
\begin{eqnarray} \label{e_queue_def}
\hspace{-200pt}{E_n}\left( {t + 1} \right) = {E_n}\left( t \right) + {\mathbf1_{n \in {\mathcal{N}_H} \cup {\mathcal{N}_M}}}{e_n}\left( t \right) \nonumber \\
\quad\quad + {\mathbf1_{n \in {\mathcal{N}_G} \cup {\mathcal{N}_M}}}{g_n}\left( t \right) - p_n^{Total}\left( t \right)
\end{eqnarray}
Where ${\mathbf1_{n \in {\mathcal{N}_H} \cup {\mathcal{N}_M}}}$ and ${\mathbf1_{n \in {\mathcal{N}_G} \cup {\mathcal{N}_M}}}$ are indicator functions. At any slot $t$, the total energy consumption must satisfy the following energy-availability constraint:
\begin{equation} \label{e_p_constraint}
{E_n}\left( t \right) \geqslant p_n^{Total}\left( t \right),\quad \forall n \in \mathcal{N}
\end{equation}
Suppose the batteries have the limited capacity $\theta _n^E$. So we have
\begin{equation} \label{e_theta_constraint}
{E_n}\left( t \right) + {\mathbf1_{n \in {\mathcal{N}_H} \cup {\mathcal{N}_M}}}{e_n}\left( t \right) + {\mathbf1_{n \in {\mathcal{N}_G} \cup {\mathcal{N}_M}}}{g_n}\left( t \right) \leqslant \theta _n^E
\end{equation}
The energy acquiring ${e_n}\left( t \right)$ and ${g_n}\left( t \right)$ should satisfy the constraint $0 \leqslant {e_n}\left( t \right) \leqslant {h_n}\left( t \right)$ and $0 \leqslant {g_n}\left( t \right) \leqslant g_n^{\max }$, respectively.
${h_n}\left( t \right)$ represents the available amount of harvesting energy at slot $t$, which should satisfy the condition
$0 \leqslant {h_n}\left( t \right) \leqslant {h_{\max }}$. Let ${\mathbf S^H}\left( t \right) = \left( {{h_n}\left( t \right),n \in {\mathcal{N}_H} \cup {\mathcal{N}_M}} \right)$ denote the harvestable energy state vector, which  is randomly varying over time slots in an i.i.d. fashion according
to a potentially unknown distribution and taking non-negative values from a
finite but arbitrarily large set ${\cal S}^H$.

\subsection{Electricity Price}
We denote the cost of per unit electricity as $p_n^G\left( t \right)$. In general, $p_n^G\left( t \right)$ depends on the electricity drawn from the electricity grid ${g_n}\left( t \right)$ and electricity price state $S_n^G\left( t \right)$. It means that $p_n^G\left( t \right)$ will change over time and space. We assume that is a stationary process with i.i.d.. Assume that $S_n^G\left( t \right)$ takes non-negative values from a finite but arbitrarily large set $\mathbf S^G$. Denote the price state vector as $\mathbf S^G = \left\{ {S_n^G\left( t \right),n \in {\mathcal{N}_G} \cup\mathcal{N}_M} \right\}$. Then we can give the price function as:
\[p_n^G\left( t \right) = p_n^G\left( {S_n^G\left( t \right),{g_n}\left( t \right)} \right)\]
For each given $S_n^G\left( t \right)$, $p_n^G\left( t \right)$ is assumed to be a increasing and continuous convex function of ${g_n}\left( t \right)$.

\subsection{Optimization Problem}

Let $U_n^f\left( x \right)$ be a continuous, concave, and non-decreasing utility function with $U_n^f\left( 0 \right) = 0$, $n \in \mathcal N$, $f \in \mathcal F$. Assume that $\beta_n^f $ is the maximum right-derivative of $U_n^f\left( x \right)$, and $0< \beta_n^f  < \infty $. We use schedule algorithms that stabilize all the queues in the system. Then for each session $f \in \mathcal{F}_n$ with source node $n \in {\mathcal{N}_s}$, we have the following condition according to the Rate Stability Theorem, i.e., Theorem 2.4 in \cite{NeelyBook2010}:
\[\bar r_n^f \leqslant \bar \mu _{n\cdot}^f + \bar d_n^f\quad n \in {\mathcal{N}_s},f \in {\mathcal{F}_n}\]
Where $\bar r_n^f$ is the time average rate of accepting packets, $\bar \mu _{n\cdot}^f$ is the time average rate of total served packets and $\bar d_n^f$ denotes the time average amount of dropping packets for session $f$ at node $n$.
As a result, we can use the value of $\bar r_n^f - \bar d_n^f$ to denote the throughput of the source node $n$ for session $f$. And we desire a solution to the following problem:
\begin{eqnarray}\label{objective_1}
  Maximize:&& \sum\limits_{n \in {\mathcal{N}_s}} {\sum\limits_{f \in {\mathcal{F}_n}} {U_n^f\left( {\bar r_n^f - \bar d_n^f} \right)} }   \\
  Subject\;to: && {\text{all queues }}Q_n^f\left( t \right){\text{ are mean rate stable}}  \nonumber
\end{eqnarray}
Notice that it will be puzzled if Lyapunov optimization is directly used to solve problem \eqref{objective_1}.

Considering the characteristic of the function $U_n^f\left( x \right)$, the following inequality should be satisfied,
\[U_n^f\left( {\bar r_n^f - \bar d_n^f} \right) \geqslant U_n^f\left( {\bar r_n^f} \right) - \beta_n^f \bar d_n^f\]
Then, let us consider the problem \eqref{objective_2}
\begin{eqnarray}\label{objective_2}
  Maximize:&& \sum\limits_{n \in {\mathcal{N}_s}} {\sum\limits_{f \in {\mathcal{F}_n}} {U_n^f\left( {\bar r_n^f} \right)} }  - \sum\limits_{n,f} {{\beta _n^f}\bar d_n^f}   \\
  Subject\;to:&&{\text{all queues }}Q_n^f\left( t \right){\text{ are mean rate stable}} \nonumber
\end{eqnarray}
We can see that the problem \eqref{objective_2} is to maximize the utility of average throughput while minimizing the amount of average drop packets as much as possible. If we can transform the problem \eqref{objective_1} to \eqref{objective_2}, the problem can be solved simply by applying Lyapunov optimization. \cite{NeelyBook2010} provides a method that completes the transform successfully.

According to the Jensen's inequality for concave functions which states that:
\[\mathbb{E}\left\{ {U_n^f\left( \tilde r_n^f\left( t \right)  \right)} \right\} \leqslant U_n^f\left( {\mathbb{E}\left\{ \tilde r_n^f \left( t \right) \right\}} \right),\; \mathbb{E}\left\{ \tilde r_n^f\left( t \right)  \right\} \in \mathbb{R}\]
We can get that the condition below would be satisfied for all $t > 0$:
\begin{eqnarray*}
\frac{1}{t}\sum\limits_{\tau  = 0}^{t - 1}  {U_n^f\left( \tilde r_n^f\left( \tau  \right) \right)}   \leqslant U_n^f\left( {\frac{1}{t}\sum\limits_{\tau  = 0}^{t - 1}  \tilde r_n^f\left( \tau  \right)  } \right),
\frac{1}{t}\sum\limits_{\tau  = 0}^{t - 1}  \tilde r_n^f\left( \tau  \right) \in \mathbb{R} \\
\text{and   } \frac{1}{t}\sum\limits_{\tau  = 0}^{t - 1} {\mathbb{E}\left\{ {U_n^f\left( \tilde r_n^f\left( \tau  \right) \right)} \right\}}  \leqslant U_n^f\left( {\frac{1}{t}\sum\limits_{\tau  = 0}^{t - 1} {\mathbb{E}\left\{ \tilde r_n^f\left( \tau  \right) \right\}} } \right)\\
,
\frac{1}{t}\sum\limits_{\tau  = 0}^{t - 1} {\mathbb{E}\left\{ \tilde r_n^f\left( \tau  \right) \right\}} \in \mathbb{R}
\end{eqnarray*}
Take limits as $t \to \infty $, then
\begin{equation} \label{jensen}
\overline {U_n^f\left( \tilde r_n^f \right)}  \leqslant U_n^f\left( \bar {\tilde r}_n^f \right), \bar {\tilde r}_n^f \in \mathbb{R}
\end{equation}
where $U_n^f\left( \tilde r_n^f \right) $ and $\bar {\tilde r}_n^f$ are defined as:
\begin{eqnarray*}
\overline {U_n^f\left( \tilde r_n^f \right)} &\triangleq& \lim_{t\to\infty}\frac{1}{t}\sum\limits_{\tau  = 0}^{t - 1} {\mathbb{E}\left\{ {U_n^f\left( \tilde r_n^f\left( \tau  \right) \right)} \right\}}, \\
\bar {\tilde r}_n^f &\triangleq& \lim_{t\to\infty}\frac{1}{t}\sum\limits_{\tau  = 0}^{t - 1} {\mathbb{E}\left\{ \tilde r_n^f\left( \tau  \right) \right\}}
\end{eqnarray*}

As presented in \cite{NeelyBook2010}, we need to construct a virtual queue to complete the transform. For each session $f \in {\mathcal{F}_n}$ at node $n \in { {\cal N}_s}$, we define a virtual flow state queue $Z_n^f\left( t \right)$, which has the queue dynamic as follows:
\begin{equation}\label{z_queue_def}
Z_n^f\left( {t + 1} \right) = \max \left\{ {\left( {Z_n^f\left( t \right) - r_n^f\left( t \right) + \tilde r_n^f\left( t \right)} \right),0} \right\}
\end{equation}
Where $\tilde r_n^f\left( t \right)$ is a auxiliary variable that satisfies the constraint $\tilde r_n^f\left( t \right) \leqslant {R_{\max }}$. We call it as the virtual input rate.

At last, we will take the electricity price into consideration.
The finally goal is to achieve the optimal trade-off between the time-average throughput utility of the source nodes and the time average cost of energy consumption in electricity grid. The optimization problem \textbf{P1} can be given as:
\begin{eqnarray}\label{objective_final}
  Maximize:&&{\varpi _1}\left( {\sum\limits_{n \in {\mathcal{N}_s}} {\sum\limits_{f \in {\mathcal{F}_n}} {\overline {U_n^f\left( {\tilde r_n^f\left( t \right)} \right)} } }  - \sum\limits_{n,f} {\overline{\beta _n^f D_n^f\left( t \right)}} } \right) \nonumber\\
  &&\hspace{-5mm}  - \left( 1-\varpi _1\right) \varpi _2\sum\limits_{n \in {\mathcal{N}_G} \cup {\mathcal{N}_M}} {\overline {p_n^G\left( t \right){g_n}\left( t \right)} } \\
  Subject\;to:&& \eqref{p_t_constraint},\eqref{miu_constraint2},\eqref{e_p_constraint},\eqref{e_theta_constraint}  \nonumber\\
  &&\bar {\tilde r}_n^f \leqslant {{\bar r}_n^f}\quad n \in \mathcal{N}_s ,f \in \mathcal{F}_n \\
  &&0 \leqslant \tilde r_n^f\left( t \right) \leqslant {R_{\max }}\quad n \in \mathcal{N}_s ,f \in \mathcal{F}_n \label{r_bar_constraint}\\
  &&0 \leqslant r_n^f\left( t \right) \leqslant {R_{\max }}\quad n \in \mathcal{N}_s , f \in \mathcal{F}_n \label{r_constraint} \\
  &&0 \leqslant D_n^f\left( t \right) \leqslant {D_{\max }}\quad f \in \mathcal{F},n \in \mathcal{N} \label{d_constraint}\\
  &&0 \leqslant {e_n}\left( t \right) \leqslant {h_n}\left( t \right)\quad n \in \mathcal{N} \label{e_constraint} \\
  &&0 \leqslant {g_n}\left( t \right) \leqslant g_n^{\max }\quad n \in \mathcal{N}\label{g_constraint} \\
  &&0 \leqslant {h_n}\left( t \right) \leqslant {h_{\max }}\quad n \in \mathcal{N} \label{h_constraint}\\
  &&\text{all queues }Q_n^f\left( t \right),\;\tilde Q_n^f\left( t \right),\;E_n\left( t \right)\text{, }Z_n^f\left( t \right) \nonumber \\
  &&\text{are mean rate stable with queuing dynamics}\nonumber \\
  && \eqref{data_q_def}, \eqref{qbar_queue_def_1}, \eqref{qbar_queue_def_2}, \eqref{e_queue_def} \text{ and } \eqref{z_queue_def}
  \text{ for }\forall n \in \mathcal{N} , \nonumber \\
  &&\forall f \in \mathcal{F} \text{, respectively.}
\end{eqnarray}
$\varpi _1$ is a weight parameter to combine these two objective functions together into a single one. $\varpi _2$ is a mapping parameter to ensure the objective functions at the same level.

\section{Cross-Layer Control via Lyapunov Optimization}
Now we will apply the Lyapunov optimization algorithm to solve the problem \textbf{P1}. First, define the network state vector at time slot $t$ as
$\mathbf\Psi  \left( t \right) \triangleq \left[\mathbf Q\left( t \right),\mathbf {\tilde Q}\left( t \right),\mathbf Z\left( t \right),\mathbf E\left( t \right) \right]$
and define the Lyapunov function $L\left( {\mathbf\Psi \left( t \right)} \right)$ by:
\begin{eqnarray}\label{L}
L\left( {\mathbf\Psi \left( t \right)} \right) &=& \frac{1}{2}\sum\limits_{n \in \mathcal{N}} {\sum\limits_{f \in \mathcal{F}} {\left[ {{{\left( {Q_n^f\left( t \right)} \right)}^2}+ {{\left( {\tilde Q_n^f\left( t \right)} \right)}^2}} \right]} } \nonumber  \\
&+& \frac{1}{2}\sum\limits_{f \in \mathcal{F}} {{{\left( {Z_n^f\left( t \right)} \right)}^2}}
 + \frac{1}{2}\sum\limits_{n \in \mathcal{N}} {{{\left( {E_n^{}\left( t \right) - \theta _n^E} \right)}^2}}
\end{eqnarray}
So the conditional Lyapunov drift at time slot $t$ can be given by:
\begin{equation}\label{drift}
\Delta \left( {\mathbf\Psi \left( t \right)} \right) = \mathbb{E}\left\{ {L\left( {\mathbf\Psi \left( {t + 1} \right)} \right) - L\left( {\mathbf\Psi \left( t \right)} \right)|\mathbf\Psi \left( t \right)} \right\}
\end{equation}
At last, we can define the drift-plus-penalty function as:
\begin{equation}\label{drift_p}
{\Delta _V}\left( {\mathbf\Psi \left( t \right)} \right) = \Delta \left( {\mathbf\Psi \left( t \right)} \right) - V\mathbb{E}\left\{ {\phi \left( t \right)|\mathbf\Psi \left( t \right)} \right\}
\end{equation}
Where
\begin{eqnarray} \label{fi}
\phi \left( t \right) &=& {\varpi _1}\left( {\sum\limits_{n \in {\mathcal{N}_s}} {\sum\limits_{f \in {\mathcal{F}_n}} { {U_n^f\left( {\tilde r_n^f\left( t \right) } \right)} } }  - \sum\limits_{n,f} {{\beta _n^f} D_n^f\left( t \right) } } \right) \nonumber\\
  &&\hspace{-5mm}  - \left( 1-\varpi _1\right) \varpi _2\sum\limits_{n \in {\mathcal{N}_G} \cup {\mathcal{N}_M}}  {p_n^G\left( t \right){g_n}\left( t \right)}
\end{eqnarray}
So we can find the drift-plus-penalty satisfied the inequality as \eqref{upperbound1}. Taking expectation on both sides of the inequality and combining \eqref{p_consumption_def} with \eqref{upperbound1}, we can transform \eqref{upperbound1} into \eqref{upperbound2}.
\begin{figure*}
\begin{eqnarray}
  {\Delta _V}\left( {\mathbf\Psi \left( t \right)} \right) \leqslant &B& - \mathbb{E}\left\{ V{\varpi _1}\left( {\sum\limits_{n \in {\mathcal{N}_s}} {\sum\limits_{f \in {\mathcal{F}_n}} { {U_n^f\left( {\tilde r_n^f\left( t \right)} \right)} } }  - \sum\limits_{n,f} {{\beta _n^f} D_n^f\left( t \right)} } \right)- V \left(1-\varpi _1\right)\varpi _2\sum\limits_{n \in {\mathcal{N}_G} \cup {\mathcal{N}_M}}  {p_n^G\left( t \right){g_n}\left( t \right)}    |\mathbf\Psi \left( t \right) \right\} \nonumber \\
  &+& \sum\limits_{n \in \mathcal{N}} {\sum\limits_{f \in \mathcal{F}} {Q_n^f\left( t \right)\mathbb{E}\left\{ {\sum\limits_{a \in \mathcal I\left( n \right)} {\mu _{an}^f\left( t \right)}  + \mathbf 1_n^fr_n^f\left( t \right) - \sum\limits_{b \in \mathcal {O}\left( n \right)} {\mu _{nb}^f\left( t \right)}  - D_n^f\left( t \right)|\mathbf\Psi \left( t \right)} \right\}} } \nonumber \\
  &+& \sum\limits_{n \in \mathcal{N}} {\sum\limits_{f \in \mathcal{F}} {\tilde Q_n^f\left( t \right)\mathbb{E}\left\{ {\varepsilon _n^f - \sum\limits_{b \in \mathcal {O}\left( n \right)} {\mu _{nb}^f\left( t \right)}  - D_n^f\left( t \right)|\mathbf\Psi \left( t \right)} \right\}} }
  + \sum\limits_{n \in {\mathcal{N}_s}} {\sum\limits_{f \in {\mathcal{F}_n}} {Z_n^f\left( t \right)\mathbb{E}\left\{ {\tilde r_n^f\left( t \right) - r_n^f\left( t \right)|\mathbf\Psi \left( t \right)} \right\}} } \nonumber  \\
  &+& \sum\limits_{n \in \mathcal{N}} {\left( {{E_n}\left( t \right) - \theta _n^E} \right)\mathbb{E}\left\{ {\mathbf 1_{n \in {{\mathcal{N}}_H} \cup {{\mathcal{N}}_M}} \cdot {e_n}\left( t \right) + {\mathbf 1_{n \in {{\mathcal{N}}_G} \cup {{\mathcal{N}}_M}}} \cdot {g_n}\left( t \right) - p_n^{Total}\left( t \right)|\mathbf\Psi \left( t \right)} \right\}}
\label{upperbound1}
\end{eqnarray}
\hrule
\end{figure*}
\begin{figure*}
\begin{eqnarray}
  \mathbb{E}\left\{ \Delta _V \left( \mathbf\Psi \left( t \right) \right) \right\} \leqslant &B& + \sum\limits_{n \in \mathcal N_s} \sum\limits_{f \in \mathcal{F}_n} \left[ Q_n^f\left( t \right) - Z_n^f \left( t \right) - \left( E_n \left( t \right) - \theta _n^E \right)\tilde p_f^S  \right]  \cdot {r_n^f}\left( t \right) \nonumber \\
  &+& \sum\limits_{n \in {\mathcal{N}_s}} \sum\limits_{f \in {\mathcal{F}_n}} {\left[ {{Z_n^f}\left( t \right) \cdot {{\tilde r}_n^f}\left( t \right) - V{\varpi _1}U_n^f\left( {{{\tilde r}_n^f}\left( t \right)} \right)} \right]}
  - \sum\limits_{n \in \mathcal{N}} {\sum\limits_{f \in \mathcal{F}} {\left[ {\tilde Q_n^f\left( t \right) + Q_n^f\left( t \right) - V{\varpi _1}\beta_n^f } \right]} }  \cdot D_n^f\left( t \right)  \nonumber \\
  &+& \sum\limits_{n \in {{\mathcal{N}}_H} \cup {{\mathcal{N}}_M}} {\left( {{E_n}\left( t \right) - \theta _n^E} \right) \cdot {e_n}\left( t \right)}  \nonumber \\
  &+& \sum\limits_{n \in {{\mathcal{N}}_G} \cup {{\mathcal{N}}_M}} {\left[ {\left( {{E_n}\left( t \right) - \theta _n^E} \right) + V\left(1-\varpi _1\right) \varpi _2 S_n^G\left( t \right)} \right] \cdot {g_n}\left( t \right)} \nonumber  \\
  &-& \sum\limits_{n \in \mathcal{N}} {\sum\limits_{f \in \mathcal{F}} {\sum\limits_{b \in {\mathcal {O}}\left( n \right)} {\left[ {Q_n^f\left( t \right) - Q_b^f\left( t \right) + \left( {{E_b}\left( t \right) - \theta _b^E} \right)\tilde p_b^R + \tilde Q_n^f\left( t \right)}  \right]\mu _{nb}^f\left( t \right)} } }  \nonumber \\
  &-& \sum\limits_{n \in \mathcal{N}} {\sum\limits_{b \in {\mathcal {O}}\left( n \right)} {\left( {{E_n}\left( t \right) - \theta _n^E} \right)p_{nb}^T\left( t \right)} }
  + \sum\limits_{n \in \mathcal{N}} {\sum\limits_{f \in \mathcal{F}} {\tilde Q_n^f\left( t \right)\varepsilon _n^f} }
\label{upperbound2}
\end{eqnarray}
\hrule
\end{figure*}

Where $B$ is a constant and satisfies:
\begin{eqnarray}\label{b}
  B &\geqslant& \frac{1}{2}\sum\limits_{n \in {\mathcal{N}}} {\sum\limits_{f \in {\mathcal{F}}} {{{\left[ {\sum\limits_{b \in {\mathcal{O}}\left( n \right)} {\mu _{nb}^f\left( t \right)}  + D_n^f\left( t \right) - \sum\limits_{a \in {\mathcal{I}}\left( n \right)} {\mu _{an}^f\left( t \right)}  }\right.} }} }\nonumber \\
  &-& { \mathbf 1_n^fr_n^f\left( t \right)} \Bigg]^2
  + \frac{1}{2}\sum\limits_{f \in {\mathcal{F}}} {{{\left[ {{{\tilde r}_n^f}\left( t \right) - {r_n^f}\left( t \right)} \right]}^2}} \nonumber \\
  &+& \frac{1}{2}\sum\limits_{n \in {\mathcal{N}}} {\sum\limits_{f \in {\mathcal{F}}} {{{\left[ {\sum\limits_{b \in {\mathcal{O}}\left( n \right)} {\mu _{nb}^f\left( t \right)}  + D_n^f\left( t \right) - \varepsilon _n^f} \right]}^2}} }  \nonumber \\
  &+& \frac{1}{2}\sum\limits_{n \in {\mathcal{N}}} {{{\left[ {{\mathbf1_{n \in {{\mathcal{N}}_H} \cup {{\mathcal{N}}_M}}}{e_n}\left( t \right) + {\mathbf1_{n \in {{\mathcal{N}}_G} \cup {{\mathcal{N}}_M}}}{g_n}\left( t \right) }\right.}}}\nonumber\\
  &-& \left.{p_n^{Total}\left( t \right)} \right]^2
\end{eqnarray}
According to \eqref{miu_constraint2} and \eqref{r_bar_constraint}-\eqref{h_constraint}, we can see such a constant must be exist.


\subsection{Framework of CLCA}
We now present our algorithm CLCA. The main design principle
of CLCA is to minimize
the R.H.S. of (\ref{upperbound2}) subject to the constraints  \eqref{p_t_constraint},\eqref{miu_constraint2},\eqref{e_p_constraint},\eqref{e_theta_constraint},
 \eqref{r_bar_constraint}-\eqref{h_constraint}.
The framework of CLCA is described in Algorithm 1 summarized in TABLE I.

\begin{table}[htbp]
\centering
\caption{Algorithm: CLCA}
\begin{tabular}{|p{3in}|}
\hline
\begin{itemize}
\item [1:] \; Initialization:
The perturbed variables ${\bm \theta}_n^E$, persistent arrival $\varepsilon _n^f$ and the penalty parameter $V$ are given, each queue blacklog is set to zero.
\item [2:] \; Observe ${S^C}\left( t \right)$, ${S^H}\left( t \right)$, ${S^G}\left( t \right)$ while given $\mathbf\Psi \left( t \right)$(the current queue backlogs are known each slot)
\item [3:] \; Choose the optimal variables to minimize the right-hand-side (RHS) of \eqref{upperbound2} subject to the constraints  \eqref{p_t_constraint},\eqref{miu_constraint2},\eqref{e_p_constraint},\eqref{e_theta_constraint},
 \eqref{r_bar_constraint}-\eqref{h_constraint}.
\item [4:] \; Update data queues, delay queues, Z queues and  the energy queues according to \eqref{data_q_def}, \eqref{qbar_queue_def_1}, \eqref{qbar_queue_def_2}, \eqref{e_queue_def} and \eqref{z_queue_def}, respectively.
\item [5:] \; Repeat step 2 to step 4 at each time slot $t \in \cal T$.
\end{itemize}
\\
\hline
\end{tabular}
\end{table}

\textbf{Remark} Note that the algorithm CLCA only requires the knowledge
of the instant values of ${S^C}\left( t \right)$, ${S^H}\left( t \right)$, ${S^G}\left( t \right)$. It does not
require any knowledge of the statistics of these stochastic
processes. The remaining challenge is to solve the problem \textbf{P2}, which is
discussed below.

\subsection{Components of CLCA}
At each time slot $t$, after observing ${S^C}\left( t \right)$, ${S^H}\left( t \right)$, ${S^G}\left( t \right)$, all components of CLCA is iteratively implemented in the distributed manner to cooperatively solve the problem \textbf{P2}.
Next, we describe each component of CLCA in detail.

\subsubsection{Source Rate Control}

For each session $f \in {\mathcal{F}_n}$ at source node $n \in {\mathcal{N}_s}$, choose $r_n^f\left( t \right)$ to solve
\begin{eqnarray} \label{r_subproblem}
  \min_{r_n^f} && \left[ {Q_n^f\left( t \right) - {Z_n^f}\left( t \right) - \left( {{E_n}\left( t \right) - \theta _n^E} \right)\tilde p_f^S} \right] \cdot r_n^f\left( t \right)  \\
  s.t.&& 0 \leqslant r_n^f\left( t \right) \leqslant {R_{\max }} \nonumber
\end{eqnarray}
It is easy to find that we can choose $r_n^f\left( t \right)$ by
\[r_n^f\left( t \right) = \left\{ \begin{gathered}
  {R_{\max }}\quad Q_n^f\left( t \right) < {Z_n^f}\left( t \right) + \left( {{E_n}\left( t \right) - \theta _n^E} \right)\tilde p_f^S \\
  0\quad \quad\;\; Q_n^f\left( t \right) \geqslant {Z_n^f}\left( t \right) + \left( {{E_n}\left( t \right) - \theta _n^E} \right)\tilde p_f^S
\end{gathered}  \right.\]

\subsubsection{Virtual Input Rate Control}

For each $f \in {\mathcal{F}_n}$, choose $\tilde r_n^f\left( t \right)$ to solve
\begin{eqnarray}\label{auxiliary_subproblem}
  &\mathop{\min}\limits_{\tilde {r}_n^f}& \quad Z_n^f\left( t \right) \cdot \tilde r_n^f\left( t \right) - V{\varpi _1}U_n^f\left( {\tilde r_n^f\left( t \right)} \right) \\
  &s.t.&\quad \quad 0 \leqslant \tilde r_n^f\left( t \right) \leqslant {R_{\max }} \nonumber
\end{eqnarray}
which is a convex optimization problem, and thus has a global optimum.

\subsubsection{Packet Drop Decision}

For each session $f \in \mathcal{F}$ and each node $n \in \mathcal{N}$, choose $D_n^f\left( t \right)$ to solve,
\begin{eqnarray}\label{packet_drop_subproblem}
  &\mathop{\max}\limits_{D_n^f}& \quad \left( {\tilde Q_n^f\left( t \right) + Q_n^f\left( t \right) - V{\varpi _1}\beta_n^f } \right) \cdot D_n^f\left( t \right)  \\
  &s.t.&\quad \quad 0 \leqslant D_n^f\left( t \right) \leqslant {D_{\max }} \nonumber
\end{eqnarray}
We can get the solution as follows:
\[D_n^f\left( t \right) = \left\{ \begin{gathered}
  {D_{\max }}\quad \quad Q_n^f\left( t \right) + \tilde Q_n^f\left( t \right) > V{\varpi _1}\beta_n^f   \\
  0\quad \quad \quad\quad Q_n^f\left( t \right) + \tilde Q_n^f\left( t \right) \leqslant V{\varpi _1}\beta_n^f
\end{gathered}  \right.\]
\textbf{Remark} According to the solution above, we can see that there will be a frequent packet drop as the backlog increases. Due to the persistence arrival of the delay queue, the delay queue backlog will increase much fast than the data queue. If the sum of $Q_n^f\left( t \right)$ and $\tilde Q_n^f\left( t \right)$ is larger than $V{\varpi _1}\beta_n^f$, the queue begin to drop packets.
%

\subsubsection{Join Optimal Transmission Power Allocation, Routing and Scheduling}

As described in previous subsection, the transmission rate over the link is associated with the transmission power. We shall consider the two variables together. So we have the optimization problem of transmission rate and transmission power as follow,
\begin{eqnarray}
 \hspace{-10mm} &\max& \sum\limits_{n \in \mathcal{N}} \sum\limits_{b \in {\mathcal {O}}\left( n \right)}  \sum\limits_{f \in \mathcal{F}}{\omega _{nb}^f\left( t \right)\mu _{nb}^f\left( t \right) }\nonumber \\
 && + \sum\limits_{n \in \mathcal{N}} {\sum\limits_{b \in {\mathcal {O}}\left( n \right)} {\left( {{E_n}\left( t \right) - \theta _n^E} \right)p_{nb}^T\left( t \right)} }   \\
 &s.t.&   0 \leqslant \sum\limits_{f \in {\mathcal{F}}} {\mu _{nb}^f\left( t \right)} \leqslant {{\tilde C}_{nb}}\left( t \right) ,\forall n \in {\mathcal{N}}, \forall b \in {\mathcal{O}}\left( n \right)  \nonumber \\
  \hspace{-10mm}\quad \quad \quad && 0 \leqslant \sum\limits_{b \in \mathcal{O}\left( n \right)} {p_{nm}^T\left( t \right)}  \leqslant P_n^{\max },\forall n \in {\mathcal{N}},\forall b \in {\mathcal{O}}\left( n \right)  \nonumber
\end{eqnarray}
where
\begin{equation}\label{weightdefinition}
\omega _{nb}^f\left( t \right) \triangleq Q_n^f\left( t \right) - Q_b^f\left( t \right) + \left( {{E_b}\left( t \right) - \theta _b^E} \right)\tilde p_b^R + \tilde Q_n^f\left( t \right)
\end{equation}
as the weight of session $f$ over link $\left( {n,b} \right)$.

\textbf{Remark}
In traditional back-pressure algorithm,
the network stability is achieved at the expense of large
packet queue backlogs. In contrast, in our proposed algorithm CLCA, the realistic
packet queue backlogs are also shared by our proposed virtual delay queues. We assign the weight
as a sum of actual packet queue backlog differential
and the backlog of a designed virtual queue, shown in (\ref{weightdefinition}).
Thus, the network stabilization is
achieved with the help of virtual queue structures that do not
contribute to delay in the network.

\paragraph{Routing and scheduling}

Define $\omega _{nb}^{{f^*}}\left( t \right) \triangleq \mathop {\max }\limits_{f \in {\mathcal{F}}} \omega _{nb}^f\left( t \right)$ as the corresponding optimal weight of link $\left( {n,b} \right)$, then the traffic session ${f^*}$ is selected for routing over link $\left( {n,b} \right)$ when $\omega _{nb}^{{f^*}}\left( t \right) > 0$.


  That is, we will allocate all the link capacity of  $\left( {n,b} \right)$ to session $f^*$, set $\mu_{n,b}^{f^*}\left( t\right) = \tilde {C}_{nb}\left( {\mathbf{p}^T}^*, \mathbf S^C \left( t \right) \right)$, where ${\mathbf{p}^T}^*$ is the transmission powers and $\mathbf S^C  \left( t \right)$ is the current channel state.

\paragraph{Transmission power allocation}

after routing and scheduling, we will try to make decision about the transmission power. Now we will observe the current channel state ${\mathbf{S}}_C\left( t \right)$ and select the transmission powers ${p^{T*}}$ by solving the following optimization problem,
\begin{eqnarray}\label{power_subproblem_1}
 \hspace{-6mm} &\mathop{\max}\limits_{p_{nm}^T}& \sum\limits_{n \in \mathcal{N}} {\sum\limits_{b \in {\mathcal {O}}\left( n \right)} {\left[ {\omega _{nb}^{{f^*}}\left( t \right){{\tilde C}_{nb}}\left( t \right) + \left( {{E_n}\left( t \right) - \theta _n^E} \right)p_{nb}^T\left( t \right)} \right]} }\\
 \hspace{-6mm} &s.t.&\quad \quad \quad 0 \leqslant \sum\limits_{b \in \mathcal{O}\left( n \right)} {p_{nm}^T\left( t \right)}  \leqslant P_n^{\max },\forall n \in {\mathcal{N}} \nonumber
\end{eqnarray}
To solve the problem \eqref{power_subproblem_1}, we develop a variable $\hat p_{nm}^T\left( t \right) = \log \left( {p_{nm}^T\left( t \right)} \right)$, and take logarithm of both sides of the constraint in problem \eqref{power_subproblem_1}, then the problem can be equivalently transformed into
\begin{eqnarray} \label{power_subproblem_2}
  &\mathop{\max}\limits_{\hat p_{nm}^T}& \quad \sum\limits_{n \in \mathcal{N}} {\sum\limits_{b \in {\mathcal {O}}\left( n \right)} {\left[ {\omega _{nb}^{{f^*}}\left( t \right){\Psi _{nb}}\left( {\hat p_{nm}^T\left( t \right)} \right)} \right.} } \nonumber \\
  &&\qquad \qquad\qquad\qquad \left. { + \left( {{E_n}\left( t \right) - \theta _n^E} \right){e^{\hat p_{nm}^T\left( t \right)}}} \right]  \\
  &s.t.&\quad  \log \left( {\sum\limits_{b \in \mathcal{O}\left( n \right)} {{e^{\hat p_{nm}^T\left( t \right)}}} } \right) - \log \left( {P_n^{\max }} \right) \leqslant 0,\forall n \in {\mathcal{N}} \nonumber
\end{eqnarray}
Where ${\Psi _{nb}}\left( {\hat p_{nm}^T\left( t \right)} \right)$ is defined as
\begin{eqnarray}
&& \hspace{-7mm} {\Psi _{nb}}\left( {\hat p_{}^T\left( t \right)} \right) \triangleq \log \left( {{\gamma _{nb}}\left( t \right)} \right) = \log S_{nb}^C + \hat p_{nb}^T\left( t \right) \nonumber \\
 &&  \hspace{-3mm} - \log \left( {N_0^b + \sum\limits_{a \in {{\mathcal{J}}_{n,b}}} {\sum\limits_{\left( {a,m} \right) \in {\mathcal{L}}} {\exp \left( {\log S_{ab}^C + \hat p_{am}^T\left( t \right)} \right)} } } \right)
\end{eqnarray}
We can see ${\Psi _{nb}}\left( {\hat p_{}^T} \right)$ is a strictly concave function of a logarithmically transformed power vector
$\hat p_{}^T\left( t \right)$. Due to \eqref{e_theta_constraint}, we have ${E_n}\left( t \right) \leqslant \theta _n^E$, so $\left( {{E_n}\left( t \right) - \theta _n^E} \right){e^{\hat p_{nm}^T\left( t \right)}}$  is a strictly concave function of $\hat p_{}^T\left( t \right)$. Thus, the objective of \eqref{power_subproblem_2} is a strictly convex in $\hat p_{}^T\left( t \right)$. As is also a strictly convex in $\hat p_{}^T\left( t \right)$, the problem \eqref{power_subproblem_2} is a strictly convex optimal problem, which has the global optimum.

Now we propose a distributed iterative algorithm base on block coordinate descent (BCD) method to solve the problem \eqref{power_subproblem_2} distributively. We assume that a single block of variables is optimized while the remaining blocks are fixed at each iteration. Let ${t_i}$ denote the $i{\text{-th}}$ iteration at time slot $t$. Then for each node $n \in {\mathcal{N}}$ at iteration ${t_i}$, the blocks $\mathbf{\hat p}_n^T = \left( {\hat p_{nb}^T,b \in {\mathcal{O}}\left( n \right)} \right)$ are updated through solving the following optimization problem \eqref{power_subproblem_3} while $\mathbf{\hat p}_{ - n}^T\left( {{t_i}} \right) = \left( {\mathbf{\hat p}_1^T\left( {{t_i}} \right)\mathbf{,}...\mathbf{,\hat p}_{n - 1}^T\left( {{t_i}} \right)\mathbf{,\hat p}_{n + 1}^T\left( {{t_i}} \right)\mathbf{,}...\mathbf{,\hat p}_N^T\left( {{t_i}} \right)} \right)$ are fixed.
\begin{eqnarray} \label{power_subproblem_3}
  &\mathop {\max }\limits_{\mathbf{\hat p}_n^T}&  \sum\limits_{n \in \mathcal{N}} {\sum\limits_{b \in {\mathcal {O}}\left( n \right)} {\left[ {\omega _{nb}^{{f^*}}\left( t \right){\Psi _{nb}}\left( {\mathbf{\hat p}_n^T,\mathbf{\hat p}_{ - n}^T\left( {{t_i}} \right)} \right) }\right.}} \nonumber\\
  &&\qquad\qquad\qquad\qquad    + \left.\left( {{E_n}\left( t \right) - \theta _n^E} \right){e^{\hat p_{nm}^T\left( t \right)}} \right] \\
  &s.t.& \log \left( {\sum\limits_{b \in \mathcal{O}\left( n \right)} {{e^{\hat p_{nm}^T\left( t \right)}}} } \right) - \log \left( {P_n^{\max }} \right) \leqslant 0,\forall n \in {\mathcal{N}} \nonumber
\end{eqnarray}

\subsubsection{Energy management}

For each node $n \in {\mathcal{N}}$, we have the optimization problem of $\left( {{e_n}\left( t \right),{g_n}\left( t \right)} \right)$ as follows,
\begin{eqnarray}\label{power_management}
  &\min :& \left( {{E_n}\left( t \right) - \theta _n^E} \right){\mathbf 1_{n \in {\mathcal{N}_H} \cup {\mathcal{N}_M}}} \cdot {e_n}\left( t \right) + \left[ {\left( {{E_n}\left( t \right) - \theta _n^E} \right)}\right. \nonumber \\
  &&+ \left. {V \left(1-\varpi _1\right) \varpi _2 S_n^G\left( t \right)} \right]{\mathbf1_{n \in {\mathcal{N}_Y} \cup {\mathcal{N}_M}}} \cdot {g_n}\left( t \right)\\
  &s.t.:&0 \leqslant {e_n}\left( t \right) \leqslant {h_n}\left( t \right) \nonumber \\
  && 0 \leqslant {g_n}\left( t \right) \leqslant g_n^{\max }\nonumber  \\
  &&{\mathbf1_{n \in {\mathcal{N}_H} \cup {\mathcal{N}_M}}} \cdot {e_n}\left( t \right){\text{ + }}{\mathbf1_{n \in {\mathcal{N}_Y} \cup {\mathcal{N}_M}}} \cdot {g_n}\left( t \right)+{E_n}\left( t \right)\nonumber  \\
  && \leqslant \theta _n^E  \nonumber
\end{eqnarray}
We can see energy management is composed of energy harvesting, energy purchasing and battery charge. Since $P_n^G\left( t \right)$ is increasing and continuous convex on ${g_n}\left( t \right)$ for each $S_n^G\left( t \right)$, the problem \eqref{power_management} turns out to be a standard convex optimization problem in $\left( {{e_n}\left( t \right),{g_n}\left( t \right)} \right)$ and can be solved efficiently.

\section{Algorithm Performance}

At the beginning of  the algorithm performance analysis, we will give an assumption that there exists $\delta  > 0$ such that
\begin{equation}\label{c_bar_p}
{\tilde C_{nm}}\left( {\mathbf{p}^T\left( t \right),\mathbf S^C \left( t \right)} \right) \leqslant \delta p_{nm}^T\left( t \right),\forall n \in \mathcal N,\forall m \in {\mathcal{O}}\left( n \right)
\end{equation}
Then we can have the theorems as follows.
\newtheorem{1}[0]{Theorem}
\begin{1}
Assume $\max \left\{ {\varepsilon _n^f,\mu _{\max }^{in} + {R_{\max }}} \right\} \leqslant {D_{\max }}$ holds, where $\mu _{\max }^{in}$ denotes the maximal amount of packets that node $n$ can receive from other nodes in one slot. Then under the algorithm CLCA with any fixed parameter $V > 0$,  all queues are bounded for $t > 0$, as follows
\[{E_n}\left( t \right) \leqslant \theta _n^E, {Z_n^f}\left( t \right) \leqslant {Z_{\max }}, \tilde Q_n^f\left( t \right) \leqslant {\tilde Q_{\max }}, Q_n^f\left( t \right) \leqslant {Q_{\max }}\]
Provided that
\[{E_n}\left( 0 \right) \leqslant \theta _n^E,{Z_n^f}\left( 0 \right) \leqslant {Z_{\max }},\tilde Q_n^f\left( 0 \right) \leqslant {\tilde Q_{\max }},Q_n^f\left( 0 \right) \leqslant {Q_{\max }}\]
Where the queue bounds are given by
\begin{eqnarray}
&&\theta _n^E = 2\delta V{\varpi _1}\beta_n^f + P_{n,\max }^{Total} + \delta (\mu _{\max }^{in} +{R_{\max }} +\varepsilon _n^f) \label{e_upbound}\\
&&{Z_{\max }} = V \varpi _1 \beta_n^f + {R_{\max }}\label{z_upbound} \\
&&{\tilde Q_{\max }}= V{\varpi _1}{\beta_n^f} + \varepsilon _n^f \label{q_bar_upbound}\\
&&{Q_{\max }} =V{\varpi _1}{\beta_n^f} + \mu _{\max }^{in} + {R_{\max }} \label{q_upbound}
\end{eqnarray}
\end{1}

\textbf{Proof}: Please see Appendix B.

\newtheorem{3}[0]{Theorem}
\begin{3}
Suppose random state vector $\boldsymbol\varphi\left( t \right) = \left[ {\mathbf{S^C}\left( t \right),\mathbf{S^H}\left( t \right),\mathbf{S^G}\left( t \right)} \right]$ is i.i.d. over slots and any C-additive approximation\footnote{Please see the definition of C-additive approximation in Definition 4.7 of \cite{NeelyBook2010}.} for minimizing RHS of \eqref{upperbound2} is used such that \eqref{r_subproblem}-\eqref{power_management} hold. And ${E_n}\left( 0 \right) \leqslant \theta _n^E$ for $\forall n \in \mathcal N$, ${Z_n^f}\left( 0 \right) \leqslant {Z_{\max }}, \tilde Q_n^f\left( 0 \right) \leqslant {\tilde Q_{\max }}, Q_n^f\left( 0 \right) \leqslant {Q_{\max }}$ for $\forall f \in \mathcal F$ and $\forall n \in \mathcal N$ are satisfied.
Then the achieved utility satisfies:
\begin{equation} \label{fi_proof}
\mathop {\lim }\limits_{t \to \infty } \frac{1}{t}\sum\limits_{\tau  = 0}^{t - 1} {\mathbb{E}\left\{ {\phi \left( \tau  \right)} \right\}}  \geqslant {\phi ^*} - \left( {B + C} \right)/V
\end{equation}
Where $B$ is defined in \eqref{b}, ${\phi ^*}$ is the optimal value associated with the problem \textbf{P1}.
\end{3}

\textbf{Proof}: Please see Appendix C.

\newtheorem{2}[0]{Theorem}
\begin{2}
When node $n$  allocates nonzero power for data sensing, compression or/and transmission, we have:
\begin{equation}
{E_n}\left( t \right) \geqslant p_{n,\max }^{Total},n \in \mathcal{N}
\end{equation}
\end{2}

\textbf{Proof}: Please see Appendix D.

\section{Performance Evaluation}

\begin{figure}
\centering
\includegraphics[scale=0.4,bb=50 350 520 800]{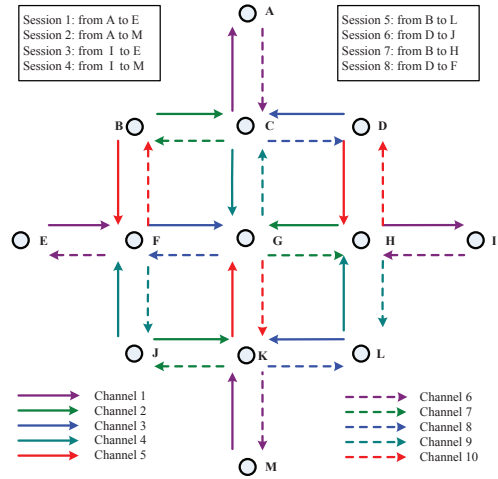}
\centering
\caption{Network topology.}
\centering
\label{fig:Topology}
\centering
\end{figure}

In this section, we will present the simulation results for our proposed algorithm CLCA.

\subsection{Simulation setting}
At first, we give the network topology as presented in Fig.\ref{fig:Topology}. In the topology, we consider a multi-channel WSNs with 13 nodes, 32 links, 8 flows/sessions transmitted on 10 different channels. We set ${\mathcal{N}_H} = \left\{ {A,C,E,H,J} \right\}$, ${\mathcal{N}_G} = \left\{ {B,F,G,L,M} \right\}$, ${\mathcal{N}_M} = \left\{ {D,I,K} \right\}$ as the default scenario.

The channel state matrix ${S^C}\left( t \right)$ hat has independent entries for every link are uniformly distributed with interval $\left[ {S_{\min }^C,S_{\max }^C} \right] \times {d^{ - 4}}$, where $S_{\min }^C = 0.9$, $S_{\max }^C = 1.1$, $d$ denotes the distance between transmitter and receiver of the link. The energy-harvesting vector ${S^H}\left( t \right)$ and the electricity price vector ${S^G}\left( t \right)$ both have independent entries that are uniformly distributed in $\left[ {0,{h_{\max }}} \right]$ and $\left[ {S_{\min }^G,S_{\max }^G} \right]$ respectively, where ${h_{\max }} = 2$, $S_{\min }^G = 0.5$ and $S_{\max }^G = 1$. All statistics of ${S^C}\left( t \right)$, ${S^H}\left( t \right)$ and ${S^G}\left( t \right)$ are i.i.d. across time-slots.

Furthermore, we set the electricity cost function as $P_n^G\left( t \right) = S_n^G\left( t \right)$ and all the initial queue size to be zero and several default values as follows:${\varpi _1} = 0.5$, ${\varpi _2} = 1$, $\delta  = 2$, $\rho = 3$, $N_0^b = 5 \times {10^{ - 13}}$, $R_{\max }^{} = 3$, ${\mu _{\max }} = 1.5$, ${D_{\max }} = 9$, $\varepsilon _n^f = 6$, $\beta _n^f = 1$, $\forall n \in \mathcal{N},\forall f \in \mathcal{F}$, $\hat p_f^S = 0.1,\forall f \in \mathcal{F}$, $P_n^{\max } = 2$, $\hat p_n^R = 0.05,\forall n \in \mathcal{N}$, $g_n^{\max } = 2,\forall n \in {N_G} \cup {N_M}$.
According to \eqref{Optimal_rho}, we set the optimal value of $\rho^*$ as XXXX to achieve the minimal worst case delay.
we set $V = \left[ {50\;150\;350\;750\;1200\;2000\;3500\;6000} \right]$. In all simulations, the simulation time is set to be $3 \times 10^5$ time slots.

\subsection{Verification of theoretic claims}
\begin{figure}
\centering
\includegraphics[scale=0.5,width=0.5\textwidth]{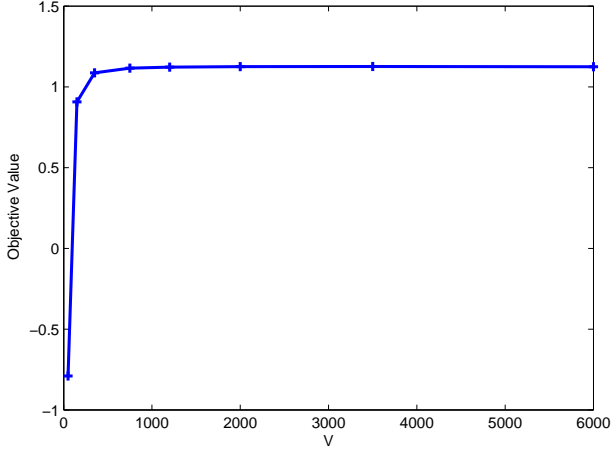}
\centering
\caption{Objective value achieved by CLCA.}
\centering
\label{fig:utility}
\centering
\end{figure}

Fig.\ref{fig:utility} presents the result of the objective value achieved by CLCA with different values of $V$. Our objective is composed of by two parts, i.e., the throughput utility and the electricity cost. Fig.\ref{fig:utility} shows that the objective value is increased with an increasing $V$ and is arbitrary close to optimal value when $V$ is large enough. This confirms the results of Theorem 3.

\begin{figure}
\centering
\includegraphics[scale=0.5,width=0.5\textwidth]{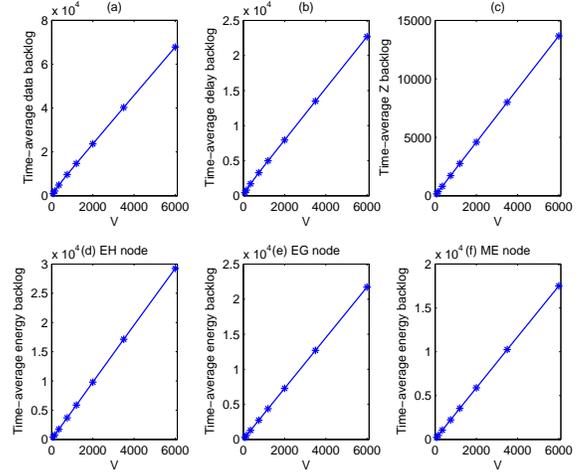}
\caption{Time average of total queue backlog in the network versus V.}
\label{fig:queue_bound}
\end{figure}

Fig. \ref{fig:queue_bound} presents the total time-average backlog in the network for six kinds of queues. From Fig. \ref{fig:queue_bound}, we observe that all queue backlogs linearly increases with increasing $V$. This shows a good match between the simulations and the claims of Theorem 1 and Theorem 2.


\subsection{Performance comparison}
Also, we provide comparisons with the existing method proposed in \cite{Neely_opportunistic2011} with the different virtual delay queue structure for achieving the worst case delay guarantees in single-hop and multi-hop networks.
Here, the virtual delay queue structure, i.e., Eq. (3) in \cite{Neely_opportunistic2011}  is applied in
the the scenario discussed in this paper.
Throughout the section and in plots, we refer to the method using the virtual queue structure proposed in \cite{Neely_opportunistic2011} by $Neely Opportunistic$.

\begin{figure}
\centering
\includegraphics[scale=0.5,width=0.5\textwidth]{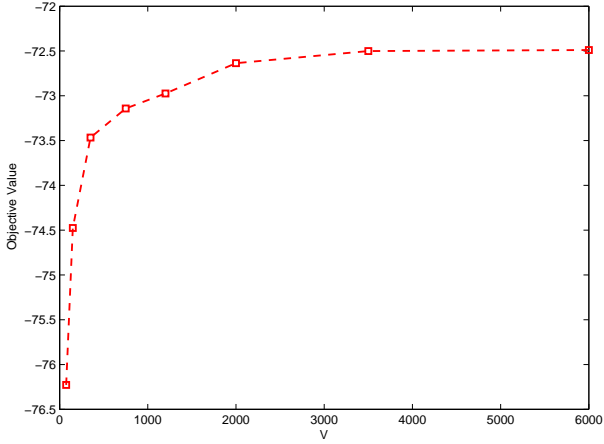}
\centering
\caption{Objective value achieved by $Neely Opportunistic$.}
\centering
\label{fig:neely_utility}
\centering
\end{figure}

\begin{figure}
\centering
\includegraphics[scale=0.5,width=0.5\textwidth]{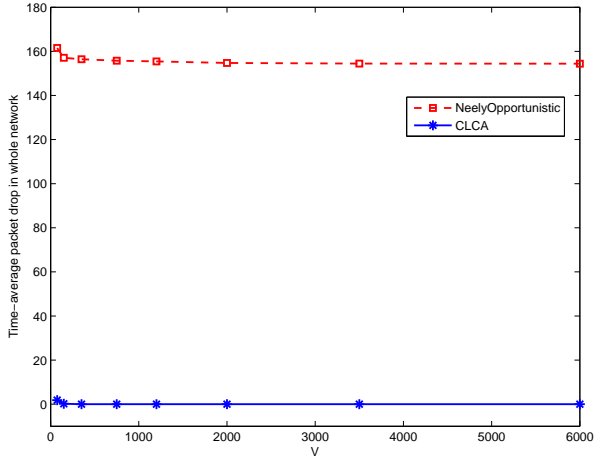}
\centering
\caption{Time-average number of dropped packets versus V.}
\centering
\label{fig:time_average_drop}
\centering
\end{figure}

In Fig. (\ref{fig:neely_utility}), we plot  the result of the objective value achieved by $Neely Opportunistic$ with different values of $V$.
It is seen that the performance of $Neely Opportunistic$ is much inferior to that of our proposed Algorithm CLCA.
The reasons is that  the virtual queue structure proposed in \cite{Neely_opportunistic2011} brings about the serious packet drop, and leads to the ultra-low throughput.
 In contrast, the virtual queue structure, Eq. \eqref{qbar_queue_def_1}\eqref{qbar_queue_def_2} used in our proposed Algorithm CLCA can achieve the zero packet drop in most case, shown in Fig. (\ref{fig:time_average_drop}).

\begin{figure}
\centering
\includegraphics[scale=0.5,width=0.5\textwidth]{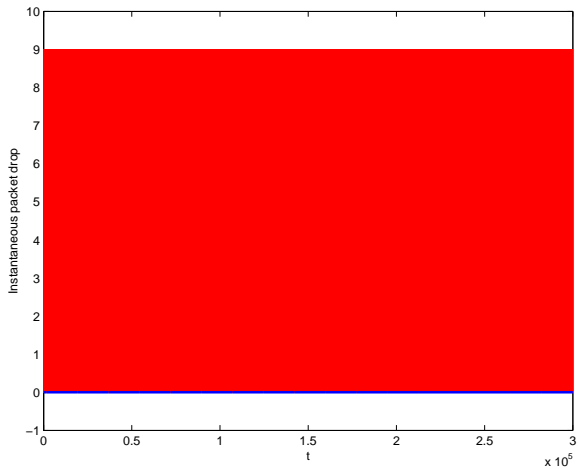}
\centering
\caption{Detailed situation of dropped packets at V=750.}
\centering
\label{fig:drop_compare}
\centering
\end{figure}

We further give the detailed situation of the packet drop of session 1 in node $A$ at $V = 750$ in Fig.\ref{fig:drop_compare}. As the green line presented in Fig.\ref{fig:drop_compare}, there is no packet drop in our proposed algorithm CLCA.  In contrast,
as the red line presented in Fig.\ref{fig:drop_compare}, there is ultra-highly frequent packet drop in $Neely Opportunistic$.

\section{Conclusions}

In this paper, we will address cross-layer control to guarantee worse case delay for Heterogeneous Powered (HP) WSNs.
We design a novel
virtual delay queue structure, and apply the Lyapunov optimization
technique to develop a cross-layer control algorithm CLCA for
HPWSNs that: (1) provide efficient throughput-utility,
(2) guarantee bounded worst-case delay, and (3) are robust to general time-varying conditions.
We develop a novel virtual delay queue scheme to share the burden of actual
packet queue backlogs to guarantee specific delay performances and finite data buffer sizes.
We analyze the performance of the proposed algorithm and verify the theoretic claims
through the simulation results.
Compared to the existing works, the algorithm presented in this
paper achieve much higher optimal objective without data drop through developing novel virtual queue structure.

\appendices

\section{Proof of Theorem 1}

Assume that for any slot $\tau  \in \left\{ {t,...,t + W_n^f + 1} \right\}$, the condition $Q_n^f\left( t \right) > \rho \tilde Q_n^f\left( t \right)$ will always be satisfied, take the beginning slot $\tau  = t + 1$, then by \eqref{qbar_queue_def_1}, we have
\begin{eqnarray*}
{\tilde Q}_n^f\left( {t + W_n^f + 1} \right) &=& {\tilde Q}_n^f\left( {t + 1} \right) \\
&&\hspace{-20mm}- \sum\limits_{\tau  = t + 1}^{\tau  = t + W_n^f} {\left( {\sum\limits_{b \in {\mathcal O}\left( n \right)} {\mu _{n{\text{b}}}^f\left( \tau  \right)}  + D_n^f\left( \tau  \right)} \right)}  + W_n^f\varepsilon _n^f
\end{eqnarray*}
Owing to $\tilde Q_n^f\left( t \right) \leqslant \tilde Q_{\max}$, then we have
\begin{eqnarray} \label{delay_1_a}
W_n^f\varepsilon _n^f \leqslant \sum\limits_{\tau  = t + 1}^{\tau  = t + W_n^f} {\left( {\sum\limits_{b \in {\mathcal O}\left( n \right)} {\mu _{n{\text{b}}}^f\left( \tau  \right)}  + D_n^f\left( \tau  \right)} \right)} + \tilde Q_{\max}\nonumber \\
 + \tilde Q_n^f\left( {t + W_n^f + 1} \right)
\end{eqnarray}
Suppose that the following inequality is satisfied,
\[\sum\limits_{\tau = t  + 1}^{\tau = t  + W_n^f} {\left( {\sum\limits_{b \in {\mathcal O}\left( n \right)} {\mu _{n{\text{b}}}^f\left( \tau  \right)}  + D_n^f\left( \tau  \right)} \right)}  \geqslant Q_n^f\left( {t + 1} \right)\]
it means that the end of the backlog $Q_n^f\left( {t + 1} \right)$ have been cleared during the interval $\tau  \in \left\{ {t + 1,...,t + W_n^f} \right\}$, and this goes against with our suppose that the worst case delay is $W_n^f$. So we can get
\begin{equation}\label{delay_1_b}
\sum\limits_{\tau = t  + 1}^{\tau = t  + W_n^f} {\left( {\sum\limits_{b \in {\mathcal O}\left( n \right)} {\mu _{n{\text{b}}}^f\left( \tau  \right)}  + D_n^f\left( \tau  \right)} \right)}  < Q_n^f\left( {t + 1} \right)
\end{equation}
Combining \eqref{delay_1_a} and \eqref{delay_1_b},
\begin{eqnarray*}
W_n^f\varepsilon _n^f &\leqslant& Q_n^f\left( {t + 1} \right) + \tilde Q_n^f\left( {t + W_n^f + 1} \right) + \tilde Q_{\max} \\
&\leqslant& Q_n^f\left( {t + 1} \right) +  Q_n^f\left( {t + W_n^f + 1} \right)/\rho + \tilde Q_{\max}\\
&\leqslant& \left[\left( 1+ \rho\right)Q_{\max }+ \rho\tilde Q_{\max}\right]/\rho
\end{eqnarray*}
Then we can get
\begin{equation}\label{delay_1}
W_n^f \leqslant \left[\left( 1+ \rho\right)Q_{\max }+ \rho\tilde Q_{\max}\right]/\left(\rho\varepsilon _n^f\right)
\end{equation}

Similarly, assume that for any slot $\tau  \in \left\{ {t,...,t + W_n^f + 1} \right\}$, the condition $Q_n^f\left( t \right) \leqslant \rho\tilde Q_n^f\left( t \right)$ will always be satisfied, we have
\begin{eqnarray*}
\tilde Q_n^f\left( {t + W_n^f + 1} \right) &=& \tilde Q_n^f\left( {t + 1} \right) \\
&+& W_n^f\left( {\varepsilon _n^f - {\mu _{\max}^{out}} - D_{\max}} \right)
\end{eqnarray*}
Thus:
\begin{eqnarray*}
W_n^f\left( {\mu _{\max}^{out} + D_{\max} - \varepsilon _n^f} \right) &=& \tilde Q_n^f\left( {t + 1} \right) -
 \tilde Q_n^f\left( {t + W_n^f + 1} \right) \\
 &\leqslant& 2\tilde Q_{\max}
\end{eqnarray*}
So we can get:
\begin{equation}\label{delay_2}
W_{\max}^f \leqslant 2{\tilde Q_{\max }}/\left( {\mu _{\max }^{out} + D_{\max } - \varepsilon _n^f} \right)
\end{equation}
Combining \eqref{delay_1} and \eqref{delay_2}, the theorem is proved.

\section{Proof of Theorem 2}

We prove Theorem 2 by induction.

\textbf{Induction Basis:} At time slot 0, the beginning of data sessions, all queues are empty. Then, we have following conditions for $\forall n \in \mathcal N, \forall f \in \mathcal F$.
\begin{eqnarray*}
Z_n^f\left( 0 \right) = 0 \leqslant V \varpi _1 \beta_n^f + R_{\max } = Z_{\max }    \\
\tilde Q_n^f\left( 0 \right) = 0 \leqslant V{\varpi _1}{\beta_n^f} + \varepsilon _n^f = \tilde Q_{\max }      \\
Q_n^f\left( 0 \right) = 0 \leqslant V{\varpi _1}{\beta_n^f} + \mu _{\max }^{in} + R_{\max } = Q_{\max }
\end{eqnarray*}

\textbf{Induction Step:} Suppose that $\forall n \in \mathcal N, \forall f \in \mathcal F$, $Z_n^f\left( t \right) \leqslant Z_{\max }$, $\tilde Q_n^f\left( t \right) \leqslant \tilde Q_{\max }$, $Q_n^f\left( t \right) \leqslant Q_{\max }$. Then for any $Z_n^f\left( t \right)$, $\tilde Q_n^f\left( t \right)$ and $Q_n^f\left( t \right)$, we have the following possible cases.
\begin{itemize}
\item $0 \leqslant Z_n^f\left( t \right) \leqslant V{\varpi _1}\beta_n^f$ or $V{\varpi _1}\beta_n^f < Z_n^f\left( t \right) \leqslant Z_{\max }$;
\item $0 \leqslant \tilde Q_n^f\left( t \right) \leqslant V{\varpi _1}{\beta_n^f}$ or $V{\varpi _1}{\beta_n^f} < \tilde Q_n^f\left( t \right) \leqslant \tilde Q_{\max }$;
\item $0 \leqslant Q_n^f\left( t \right) \leqslant V{\varpi _1}{\beta_n^f}$ or $V{\varpi _1}{\beta_n^f} < Q_n^f\left( t \right) \leqslant Q_{\max }$;
\end{itemize}

\begin{itemize}
\item [-] We first analyze the size of $Z_n^f\left( t \right)$:

\begin{itemize}
\item If $0 \leqslant Z_n^f\left( t \right) \leqslant V{\varpi _1}\beta_n^f$, according to \eqref{z_queue_def} we have
\begin{eqnarray*}
  Z_n^f\left( {t + 1} \right) &\leqslant& Z_n^f\left( t \right) + R_{\max }  \\
    &\leqslant& V{\varpi _1}\beta_n^f  + R_{\max }
\end{eqnarray*}
where the first inequality is due to the fact that $Z_n^f$ can increase by at most $R_{\max }$ in one slot, and the second inequality is according to our proposed assumption.
\item If $V{\varpi _1}\beta_n^f < Z_n^f\left( t \right) \leqslant  Z_{\max }$, then
\begin{eqnarray*}
  &&V{\varpi _1} \cdot U_n^f\left( {\tilde r_n^f\left( t \right)} \right) - Z_n^f\left( t \right)\tilde r_n^f\left( t \right)\\
  &\leqslant& V{\varpi _1} \cdot U_n^f\left( 0 \right) + V{\varpi _1}\beta_n^f \tilde r_n^f\left( t \right)
  - Z_n^f\left( t \right)\tilde r_n^f\left( t \right)  \\
  &=& V{\varpi _1} \cdot U_n^f\left( 0 \right) + \left( {V{\varpi _1}\beta_n^f  - Z_n^f\left( t \right)} \right)\tilde r_n^f\left( t \right)  \\
  &\leqslant& V{\varpi _1} \cdot U_n^f\left( 0 \right)  \\
  &=& 0
\end{eqnarray*}
where the first inequality is due to the fact that  $\beta_n^f$ is the maximum derivative of the $U_n^f\left( {\tilde r_n^f\left( t \right)} \right)$ function. Then we can get that $V{\varpi _1} \cdot U_n^f\left( {\tilde r_n^f\left( t \right)} \right) - Z_n^f\left( t \right)\tilde r_n^f\left( t \right)$ is a non-positive value. According to the sub-problem \eqref{auxiliary_subproblem}, if $Z_n^f\left( t \right) > V{\varpi _1}\beta_n^f$, we will choose $\tilde r_n^f\left( t \right) = 0$. Then according to \eqref{z_queue_def}, we have $Z_n^f\left( {t + 1} \right) \leqslant Z_n^f\left( t \right)$.
\end{itemize}
So far, we prove that $Z_n^f\left( t \right) \leqslant  Z_{\max }$, $\forall n \in \mathcal N, \forall f \in \mathcal F$ for each time slot $t$.

\item [-] Next, we analyze the size of $\tilde Q_n^f\left( t \right)$:
\begin{itemize}
\item If $0 \leqslant \tilde Q_n^f\left( t \right) \leqslant V{\varpi _1}\beta_n^f$, then
\begin{eqnarray*}
  \tilde Q_n^f\left( {t + 1} \right) \leqslant \tilde Q_n^f\left( t \right) + \varepsilon _n^f
    \leqslant V{\varpi _1}{\beta_n^f} + \varepsilon _n^f
\end{eqnarray*}
The first inequality is due to the fact that the queue $\tilde Q_n^f\left( t \right)$ can increase by at most $\varepsilon _n^f$ in one slot.
\item If $V{\varpi _1}\beta_n^f \leqslant \tilde Q_n^f\left( t \right) \leqslant \tilde Q_{\max }$, we can see that $D_n^f\left( t \right) = {D_{\max }}$ according to sub-problem \eqref{packet_drop_subproblem}. Since the condition $\varepsilon _n^f \leqslant D_{\max }$ is satisfied, we have
\begin{eqnarray*}
  \tilde Q_n^f\left( {t + 1} \right) \leqslant \tilde Q_n^f\left( t \right) - D_{\max } + \varepsilon _n^f
    \leqslant \tilde Q_n^f\left( t \right)
\end{eqnarray*}
\end{itemize}
Up to now, we prove that $\tilde Q_n^f\left( t \right) \leqslant \tilde Q_{\max }$, $\forall n \in \mathcal N, \forall f \in \mathcal F$ for each time slot $t$.

\item [-] Now, we analyze the size of $Q_n^f\left( t \right)$:
\begin{itemize}
\item If $0 \leqslant Q_n^f\left( t \right) \leqslant V{\varpi _1}{\beta_n^f}$, then we can get the below inequality according to \eqref{data_q_def}:
\begin{eqnarray*}
  Q_n^f\left( {t + 1} \right) &\leqslant& Q_n^f\left( t \right) + \sum\limits_{a \in {\mathcal I}\left( n \right)} {\mu _{an}^f\left( t \right)}  + {\mathbf{1}}_n^fR_n^f\left( t \right) \\
  &\leqslant& V{\varpi _1}{\beta_n^f} + \mu _{\max }^{in} + {R_{\max }}
\end{eqnarray*}
\item If $V{\varpi _1}{\beta_n^f} < Q_n^f\left( t \right) \leqslant Q_{\max }$, we have $D_n^f\left( t \right) = {D_{\max }}$ according to sub-problem \eqref{packet_drop_subproblem}. Since the data queue can increase by at most $\mu _{\max }^{in} + {R_{\max }}$ according to \eqref{data_q_def}, we have
\begin{eqnarray*}
  Q_n^{f}\left( {t + 1} \right) &\leqslant& Q_n^f\left( t \right) - D_{\max } + \mu _{\max }^{in} + R_{\max }  \\
    &\leqslant& Q_n^f\left( t \right)
\end{eqnarray*}
\end{itemize}
Up to now, we prove that $Q_n^f\left( t \right) \leqslant Q_{\max }$, $\forall n \in \mathcal N, \forall f \in \mathcal F$ for each time slot $t$.
\end{itemize}

So we complete the proof of Theorem 2.$\Box$

\section{Proof of Theorem 3}

We denote $\alpha _{n,f}^Q$, $\alpha _{n,f}^{\tilde Q}$, $\alpha _{n,f}^Z$, $\alpha _n^E$ and $\mu _{n,f}^Q$, $\mu _{n,f}^{\tilde Q}$, $\mu _{n,f}^Z$, $\mu _n^E$ as the input and output of the queue $Q_n^f\left( t \right)$, $\tilde Q_n^f\left( t \right)$, $Z_n^f\left( t \right)$, ${E_n}\left( t \right)$ for $\forall f\in\mathcal{F},n\in\mathcal{N}$, respectively. Denote ${\mathbf{I}^{}}\left( t \right) = \left( {\mathbf{\tilde r}\left( t \right),\mathbf{r}\left( t \right),\mathbf{D}\left( t \right),\mathbf p^T \left( t \right),\boldsymbol{\mu}\left( t \right),\mathbf{e}\left( t \right),\mathbf{g}\left( t \right)} \right)$ as the vector of variables of the problem \textbf{P1}. According to the Optimality over $\omega {\text{-only}}$ policies theorem (Theorem 4.5 in \cite{NeelyBook2010}). For all $\eta  > 0$, there exists an $\varphi {\text{-only}}$ policy $\mathbf{I^*}$ that chooses $\mathbf{I^*}\left( t \right) \in \mathcal{I}_{\mathbf{\varphi }\left( t \right)}$ as a random function of random state $\boldsymbol{\varphi }\left( t \right)$, where $\mathcal{I}_{\mathbf{\varphi }\left( t \right)}$ is an abstract set that defines decision options under state $\boldsymbol{\varphi }\left( t \right)$, such that:
\begin{eqnarray}
  &&\hspace{6mm}\phi \left( {\mathbf{I^*}\left( t \right),\boldsymbol{\varphi }\left( t \right)} \right) = {\phi ^*}\label{eta_obj}  \\
  &&\hspace{-6mm} \mathbb{E}\left\{ {\alpha _{n,f}^Q\left( {\mathbf{I^*}\left( t \right),\boldsymbol{\varphi }\left( t \right)} \right)} \right\} \leqslant \mathbb{E}\left\{ {\mu _{n,f}^Q\left( {\mathbf{I^*}\left( t \right),\boldsymbol{\varphi }\left( t \right)} \right)} \right\} + \eta   \\
  &&\hspace{-6mm}\mathbb{E}\left\{ {\alpha _{n,f}^{\tilde Q}\left( {\mathbf{I^*}\left( t \right),\boldsymbol{\varphi }\left( t \right)} \right)} \right\} \leqslant \mathbb{E}\left\{ {\mu _{n,f}^{\tilde Q}\left( {\mathbf{I^*}\left( t \right),\boldsymbol{\varphi }\left( t \right)} \right)} \right\} + \eta   \\
  &&\hspace{-6mm} \mathbb{E}\left\{ {\alpha _{n,f}^Z\left( {\mathbf{I^*}\left( t \right),\boldsymbol{\varphi }\left( t \right)} \right)} \right\} \leqslant \mathbb{E}\left\{ {\mu _{n,f}^Z\left( {\mathbf{I^*}\left( t \right),\boldsymbol{\varphi }\left( t \right)} \right)} \right\} + \eta   \\
  &&\hspace{-6mm} \mathbb{E}\left\{ {\alpha _n^E\left( {\mathbf{I^*}\left( t \right),\boldsymbol{\varphi }\left( t \right)} \right)} \right\} \leqslant \mathbb{E}\left\{ {\mu _n^E\left( {\mathbf{I^*}\left( t \right),\boldsymbol{\varphi }\left( t \right)} \right)} \right\} + \eta \label{eta_energy}  \\
  &&\hspace{-6mm} 0 \leqslant \tilde r_n^{f*}\left( t \right) \leqslant {R_{\max }} \quad f \in \mathcal{F}   \\
  &&\hspace{-6mm} 0 \leqslant r_n^{f*}\left( t \right) \leqslant {R_{\max }}\quad f \in \mathcal{F}  \\
  &&\hspace{-6mm} 0 \leqslant D_n^{f*}\left( t \right) \leqslant {D_{\max }} \quad f \in \mathcal{F},n \in {\mathcal{N}}  \\
  &&\hspace{-6mm} 0 \leqslant e_n^*\left( t \right) \leqslant {h_n}\left( t \right)\quad n \in {\mathcal{N}}  \\
  &&\hspace{-6mm} 0 \leqslant g_n^*\left( t \right) \leqslant g_n^{\max } \quad n \in {\mathcal{N}}  \\
  &&\hspace{-6mm} 0 \leqslant h_n^*\left( t \right) \leqslant {h_{\max }} \quad n \in {\mathcal{N}}  \\
  &&\hspace{-6mm} \mathbf{I^*}\left( t \right) \in \mathcal{I}_{\mathbf{\varphi }\left( t \right)} \label{i_constraint}
\end{eqnarray}

The C-additive approximation ensures by \eqref{upperbound1}:
\begin{eqnarray} \label{c_upperbound}
  {\Delta _V}\left( \mathbf\Psi \left( t \right) \right) &\leqslant& B + C - V\mathbb{E}\left\{ {\phi \left( {\mathbf{I^*}\left( t \right),\boldsymbol{\varphi }\left( t \right)} \right)|\mathbf\Psi \left( t \right)} \right\} \nonumber \\
 &+& \sum\limits_{n \in \mathcal{N}} {\sum\limits_{f \in \mathcal{F}} {Q_n^f\left( t \right)\mathbb{E}\left\{ {\alpha _{n,f}^Q\left( {\mathbf{I^*}\left( t \right),\boldsymbol{\varphi }\left( t \right)} \right) } \right.} }\nonumber\\
 && - \left.{\mu _{n,f}^Q\left( {\mathbf{I^*}\left( t \right),\boldsymbol{\varphi }\left( t \right)} \right)|\mathbf\Psi\left( t \right)} \right\} \nonumber  \\
 &+& \sum\limits_{n \in \mathcal{N}} {\sum\limits_{f \in \mathcal{F}} {\tilde Q_n^f\left( t \right)\mathbb{E}\left\{ {\alpha _{n,f}^{\tilde Q}\left( {\mathbf{I^*}\left( t \right),\boldsymbol{\varphi }\left( t \right)} \right) } \right.} }\nonumber\\
 && - \left.{\mu _{n,f}^{\tilde Q}\left( {\mathbf{I^*}\left( t \right),\boldsymbol{\varphi }\left( t \right)} \right)|\mathbf\Psi \left( t \right)} \right\} \nonumber  \\
 &+&  \sum\limits_{n \in {\mathcal{N}_s}} {\sum\limits_{f \in {\mathcal{F}_n}} {Z_n^f\left( t \right)\mathbb{E}\left\{ {\alpha _{n,f}^Z\left( {\mathbf{I^*}\left( t \right),\boldsymbol{\varphi }\left( t \right)} \right)} \right.} }\nonumber \\
 && - \left.{\mu _{n,f}^Z\left( \mathbf{I^*}\left( t \right),\boldsymbol{\varphi }\left( t \right) \right)|\mathbf\Psi \left( t \right)} \right\} \nonumber  \\
 &+&  \sum\limits_{n \in \mathcal{N}} {\left( {{E_n}\left( t \right) - \theta _n^E} \right)\mathbb{E}\left\{ {\alpha _n^E\left( {\mathbf{I^*}\left( t \right),\boldsymbol{\varphi }\left( t \right)} \right) } \right.}\nonumber\\
 && - \left.{\mu _n^E\left( \mathbf{I^*}\left( t \right),\boldsymbol{\varphi }\left( t \right) \right)|\mathbf\Psi \left( t \right)} \right\}
\end{eqnarray}

Substituting the $\boldsymbol{\varphi } {\text{-only}}$ policy from \eqref{eta_obj}-\eqref{i_constraint} in RHS of the above inequality \eqref{c_upperbound} and taking $\eta  \to 0$, then
\begin{equation}\label{deta_proof}
{\Delta _V}\left( {\mathbf\Psi \left( t \right)} \right) \leqslant B + C - V{\phi ^*}
\end{equation}

Combining \eqref{deta_proof} with \eqref{L}-\eqref{fi} and using iterated expectations and telescoping sums for all $t>0$:
\[\frac{1}{t}\sum\limits_{\tau  = 0}^{t - 1} {\phi \left( \tau  \right)}  \geqslant {\phi ^*} - \left( {B + C} \right)/V - \mathbb{E}\left\{ {L\left( {\mathbf\Psi \left( 0 \right)} \right)} \right\}/\left( {Vt} \right)\]

Taking $t \to \infty $, we can get that:
\[\mathop {\lim }\limits_{t \to \infty } \frac{1}{t}\sum\limits_{\tau  = 0}^{t - 1} {\mathbb{E}\left\{ {\phi \left( \tau  \right)} \right\}}  \geqslant {\phi ^*} - \left( {B + C} \right)/V\]
So we complete the proof of Theorem 3.$\Box$

\section{Proof of Theorem 4}

According to the definition of the link capacity, we can get that the inequality \eqref{c_bar} holds.
\begin{equation} \label{c_bar}
{\tilde C_{ab}}\left( {\mathbf{p}^T\left( t \right),\mathbf S^C\left( t \right)} \right) \leqslant {\tilde C_{ab}}\left( {\mathbf{p}^{T'}\left( t \right),\mathbf S^C\left( t \right)} \right)
\end{equation}
Where ${\tilde C_{ab}}\left( {\mathbf{p}^{T'}\left( t \right),\mathbf S^C\left( t \right)} \right)$ obtained by setting $p_{nm}^T\left( t \right)$ of $\mathbf p_{}^T\left( t \right)$ to zero, $\left( {n,m} \right) \in \mathcal{L}$, $\left( {a,b} \right) \in \mathcal{L}$ and $\left( {n,m} \right) \ne \left( {a,b} \right)$.

Then we can see the weight of session f over link $\left( {n,m} \right)$ satisfied:
\begin{eqnarray} \label{link_weight}
  \omega _{nm}^f\left( t \right) &=& Q_n^f\left( t \right) - Q_m^f\left( t \right) + \left( {{E_m}\left( t \right) - \theta _m^E} \right)\tilde p_m^R + \tilde Q_n^f\left( t \right) \nonumber \\
   &\leqslant& Q_n^f\left( t \right) + \tilde Q_n^f\left( t \right) \nonumber \\
   &\leqslant& 2V{\varpi _1}{\beta_n^f} + \mu _{\max }^{in} + {R_{\max }} + \varepsilon _n^f
\end{eqnarray}

Suppose ${E_n}\left( t \right) < p_{n,\max }^{Total}$ when node $n \in \mathcal{N}$ allocates nonzero power for data sensing, compression or transmission and the power allocation control vector $\mathbf{p}^{T*}$ is the optimal solution to sub-problem \eqref{power_subproblem_1}. Without loss of generality, there should be some $p_{mn}^{T*}\left( t \right) > 0$. We can get a vector $\mathbf{p}^T$ by setting $p_{mn}^{T*}\left( t \right) = 0$.  Let $G\left( {\mathbf{p}^T\left( t \right),\mathbf S^C\left( t \right)} \right)$ denote the objective function of sub-problem \eqref{power_subproblem_1}, so we have
\begin{eqnarray}
 \hspace{-7mm} &&G\left( {\mathbf{p}^{T*}\left( t \right),\mathbf S^C\left( t \right)} \right) - G\left( {\mathbf{p}^T\left( t \right),\mathbf S^C\left( t \right)} \right) \nonumber \\
 \hspace{-7mm} &=& \sum\limits_{n \in \mathcal{N}} {\sum\limits_{b \in \mathcal{O}\left( n \right)} {\left[ {{{\tilde C}_{nb}}\left( {\mathbf{p}^{T*}\left( t \right),\mathbf S^C\left( t \right)} \right)}\right. } }\nonumber \\
 \hspace{-10mm} &&- \left.{{{\tilde C}_{nb}}\left( {\mathbf{p}^T\left( t \right),\mathbf S^C\left( t \right)} \right)} \right]\omega _{nb}^{{f^*}}\left( t \right)
  + \left( {{E_n}\left( t \right) - \theta _n^E} \right)p_{nm}^{T*}\left( t \right) \label{G_1} \\
 \hspace{-10mm} &\leqslant& {{\tilde C}_{nm}}\left( {\mathbf{p}^{T*}\left( t \right),\mathbf S^C\left( t \right)} \right)\omega _{nb}^{{f^*}}\left( t \right) + \left( {{E_n}\left( t \right) - \theta _n^E} \right)p_{nm}^{T*}\left( t \right) \label{G_2} \\
 \hspace{-7mm} &\leqslant& \delta p_{nm}^{T*}\left( t \right)\left( {2V{\varpi _1}{\beta_n^f} + \mu _{\max }^{in} + {R_{\max }} + \varepsilon _n^f} \right) \nonumber \\
 \hspace{-7mm} &&- \left( {p_{n,\max }^{Total} - \theta _n^E} \right)p_{nm}^{T*}\left( t \right)\label{G_3}  \\
 \hspace{-7mm} &=& 0 \label{G_4}
\end{eqnarray}
where \eqref{G_1} is obtained by the sub-problem \eqref{power_subproblem_1}. We can get \eqref{G_2} by ${\tilde C_{nb}}\left( {\mathbf{p}^{T*}\left( t \right),\mathbf S^C\left( t \right)} \right) \leqslant {\tilde C_{nb}}\left( {\mathbf{p}^T\left( t \right),\mathbf S^C\left( t \right)} \right),b \ne m$ according to \eqref{c_bar}. Combining the assumption ${E_n}\left( t \right) < p_{n,\max }^{Total}$ with \eqref{c_bar_p}, \eqref{link_weight} and \eqref{G_2}, we can get \eqref{G_3}. And combining \eqref{G_3} with the energy queue upperbound \eqref{e_upbound}, we can get \eqref{G_4}. So we can see that $\mathbf p^{T*}$ is not the optimal solution to \eqref{power_subproblem_1}, which is inconsistent with our assumption. Then we have ${E_n}\left( t \right) \geqslant p_{n,\max }^{Total}$.
So we complete the proof of Theorem 4.$\Box$


\begin{thebibliography}{1}

\bibitem{Gungor_Hancke2009}
V. C. Gungor and G. P. Hancke, ``Industrial wireless sensor networks:
Challenges, design principles, and technical approaches,''
\emph{IEEE Trans. Ind. Electron.}, Vol. 56, No. 10, pp. 4258-4265,
Oct. 2009.


\bibitem{Cao_Chen2010}
X. Cao, J. Chen, Y. Xiao, Y. Sun, ``Building-Environment Control With Wireless Sensor and Actuator Networks: Centralized Versus Distributed ,'' \emph{IEEE Trans. on Industrial Electronics}, Vol. 57, No. 11, pp. 3596-3605 , Nov. 2010.




\bibitem{Wu2011}
D. Wu, S. Ci, H. Luo, Y. Ye, H. Wang, ``Video Surveillance Over Wireless Sensor and Actuator Networks Using Active Cameras,'' \emph{ IEEE Transactions on Automatic Control},
Vol.	56, No. 10, pp. 2467-2472, 2011.


\bibitem{Mo2011}
Y. Mo, E. Garone, A. Casavola, B. Sinopoli, ``Stochastic Sensor Scheduling for Energy Constrained Estimation in Multi-Hop Wireless Sensor Networks,'' \emph{IEEE Transactions on Automatic Control},
Vol. 56, No. 10, pp. 2489-2495, 2011.

\bibitem{Wang2013}
X. Wang, S. Han, Y. Wu, X. Wang, ``Coverage and Energy Consumption Control in Mobile Heterogeneous Wireless Sensor Networks,'' \emph{IEEE Transactions on Automatic Control},
Vol. 58, No. 4,	pp. 975-988, 2013.


%
%

\bibitem{Sudevalayam_Survey}
S. Sudevalayam and P. Kulkarni. Energy harvesting sensor nodes:
Survey and implications. IEEE Communications Surveys and Tutorials,
vol. 13, no. 3, pp. 443-461, 2011.







\bibitem{Gong_Niu2013}
J. Gong, S. Zhou, Z. Niu, ``Optimal Power Allocation for Energy Harvesting and Power Grid Coexisting Wireless Communication Systems,'' \emph{IEEE Transactions on Communications}, Vol. 61, No.7, pp. 3040-3049, 2013.








%
%
%


\bibitem{Hariharan2013}
S. Hariharan, Z. Zheng, N. B. Shroff, ``Maximizing Information in Unreliable Sensor Networks Under Deadline and Energy Constraints,'' \emph{IEEE Transactions on Automatic Control},
Vol. 58, No. 6,	pp. 1416-1429, 2013.

\bibitem{Zheng2014}
Z. Zheng, N. B. Shroff, ``Submodular Utility Maximization for Deadline Constrained Data Collection in Sensor Networks,'' \emph{IEEE Transactions on Automatic Control},
DOI: 10.1109/TAC.2014.2321683, 2014.


\bibitem{Yoo_Shim2012}
H. Yoo, M. Shim and D. Kim, ``Dynamic duty-cycle
scheduling schemes for energy-harvesting wireless sensor networks,''
\emph{IEEE Communications Letters}, Vol. 16, No. 2, pp. 202-204, 2012.


\bibitem{Sharma2010}
V. Sharma, U. Mukherji, V. Joseph, and S. Gupta, ``Optimal energy
management policies for EH sensor nodes,`` \emph{IEEE Transactions
Wireless Communications}, Vol. 9, No. 4, pp. 1326-1336, Apr. 2010.


\bibitem{Srivastava2013}
R. Srivastava and C. E. Koksal, ``Basic performance limits and tradeoffs
in EH sensor nodes with finite data and energy storage,''
\emph{IEEE/ACM Transactions on Networking}, Vol. 21, No. 4, pp. 1049-1062, Aug.
2013.

\bibitem{ZhangTWC2013}
Y. Zhang, S. He, J. Chen, Y. Sun, X. Shen, ``Distributed Sampling Rate Control for Rechargeable Sensor Nodes with Limited Battery Capacity,'' \emph{IEEE Transactions on Wireless Communications}, Vol. 12, No. 6, pp. 3096-3106, 2013.

\bibitem{MaoTVT2014}
S. Mao, M. Cheung, V. Wong, ``Joint Energy Allocation for Sensing and Transmission in Rechargeable Wireless Sensor Networks,''
\emph{IEEE Transactions on Vehicular Technology}, Vol. 63, No. 6, pp. 2862-2875, 2014.


\bibitem{Mao2012TAC}
Z. Mao, C. E. Koksal, and N. B. Shroff, ``Near optimal power and
rate control of multi-hop sensor networks with energy replenishment:
Basic limitations with finite energy and data storage,'' \emph{IEEE Transactions on
Automatic Control}, Vol. 57, No. 4, pp. 815-829, Apr. 2012.

\bibitem{Chen2012INFOCOM}
S. Chen, P. Sinha, N. Shroff, and C. Joo, ``A simple asymptotically
optimal energy allocation and routing scheme in rechargeable sensor
networks,'' \emph{in Proc. of IEEE INFOCOM}, Orlando, FL, USA, Mar. 2012.


\bibitem{Sarkar2013}
S. Sarkar, M.H.R. Khouzani, and K. Kar, ``Optimal Routing and Scheduling in Multihop Wireless Renewable Energy Networks,''  \emph{IEEE Transactions on Automatic Control}, Vol. 58, No. 7, pp. 1792-1798, 2013.

\bibitem{Gatzianas2010}
M. Gatzianas, L. Georgiadis, and L. Tassiulas, ``Control of wireless
networks with rechargeable batteries,'' \emph{IEEE Transactions on Wireless Communications},
Vol. 9, No. 2, pp. 581-593, Feb. 2010.

\bibitem{Huang_Neely2013}
L. Huang and M. J. Neely, ``Utility optimal scheduling in Energy-Harvesting networks,''
\emph{IEEE/ACM Transactions on  Networking}, Vol. 21, No. 4, pp. 1117-1130,
 Aug. 2013.

\bibitem{Tapparello2014}
C. Tapparello, O. Simeone, M. Rossi,
``Dynamic Compression-Transmission for Energy-Harvesting Multihop Networks With Correlated Sources,'' \emph{IEEE/ACM Transactions on Networking},
DOI: 10.1109/TNET.2013.2283071, 2013.




%













%



%

%

\bibitem{survey_yingcui}
Y. Cui, V. K. N. Lau, R. Wang, H. Huang, S. Zhang. ``A Survey on Delay-Aware Resource Control for Wireless Systems-Large Deviation Theory, Stochastic Lyapunov Drift, and Distributed Stochastic Learning,'' \emph{IEEE Transactions on Information Theory}, Vol. 58, No. 3, pp. 1677-1701, Mar. 2012.

\bibitem{Gupta_Shroff_infocom2009}
G. Gupta and N. Shroff, ``Delay analysis for multi-hop wireless networks,''
\emph{in Proc. IEEE INFOCOM}, pp. 2356-2364, Apr. 2009.





\bibitem{Venkataramanan_Lin2010}
V. Venkataramanan, X. Lin, L. Ying, and S. Shakkottai, ``On scheduling
for minimizing end-to-end buffer usage over multihop wireless
networks,'' \emph{in Proc. IEEE INFOCOM}, pp. 1-9, Mar. 2010.




\bibitem{Bui_Srikant2009}
L. Bui, R. Srikant, and A. Stolyar, ``Novel architectures and algorithms
for delay reduction in back-pressure scheduling and routing,'' \emph{in Proc.
IEEE INFOCOM Mini-Conf.}, pp. 2936-2940, Apr. 2009.

\bibitem{Ying_Shakkottai2009}
L. Ying, S. Shakkottai, and A. Reddy, ``On combining shortest-path
and back-pressure routing over multihop wireless networks,'' \emph{in Proc.
IEEE INFOCOM}, pp. 1674-1682, Apr. 2009.

\bibitem{Xiong_Li2011}
H. Xiong, R. Li, A. Eryilmaz, and E. Ekici, ``Delay-aware cross-layer
design for network utility maximization in multi-hop networks,'' \emph{IEEE
J. Sel. Areas Commun.}, Vol. 29, No. 5, pp. 951-959, May. 2011.



\bibitem{Huang_Lin2013}
P. Huang, X. Lin, and C.Wang, ``A low-complexity congestion control and scheduling algorithm for multihop wireless networks with order optimal
per-flow delay,''
\emph{IEEE/ACM Transactions on Networking}, Vol. 21, No. 2, Pages. 495-508, Apr. 2013.

\bibitem{Le_Modiano2012}
L. Le, E. Modiano, and N. Shroff, ``Optimal control of wireless networks
with finite buffers,'' \emph{IEEE/ACM Trans. Netw.},  Vol. 20, No. 4, pp. 1316-1329, 2012.

\bibitem{Xue_Ekici2013}
D. Xue and E. Ekici, ``Delay-guaranteed cross-layer scheduling in multi-hop wireless networks,'' \emph{IEEE/ACM Trans. Netw.}, Vol. 21, No. 6, pp. 1696-1707, 2013.

\bibitem{Neely_opportunistic2011}
M. J. Neely, ``Opportunistic scheduling with worst case delay guarantees
in single and multi-hop networks,'' \emph{in Proc. IEEE INFOCOM}, pp. 1728-1736, Apr.
2011.







\bibitem{Neely2006Book}
L. Georgiadis, M.J. Neely, L. Tassiulas, ``Resource allocation and cross-layer control in wireless networks'', \emph{ Foundations and Trends in Networking}, vol. 1, no. 1, pp. 1-144, 2006.

\bibitem{NeelyBook2010}
M. J. Neely, ``Stochastic Network Optimization with Application to Communication and Queueing Systems,'' \emph{Synthesis Lectures on Communication Networks}, Morgan \& Claypool, 2010.


\end{thebibliography}
\end{document}